\documentclass[fleqn,10pt]{wlscirep}
\usepackage[utf8]{inputenc}
\usepackage[T1]{fontenc}
\usepackage{graphicx}	% Including figure files
\usepackage{amsmath}	% Advanced maths commands
\usepackage{amssymb}	% Extra maths symbols
\usepackage{longtable}  % Long table
\usepackage{ulem}
\usepackage{newfloat}
\DeclareFloatingEnvironment[name={Supplementary Figure},fileext=lof]{suppfigure}
\DeclareFloatingEnvironment[name={Supplementary Table},fileext=lof]{suptable}

\newcommand{\eff}{_{\textrm{\tiny eff}}} % non-italic subscript "eff"
\newcommand{\sun}{_{\odot}} % Sun-symbol
\newcommand{\WD}{_\textrm{WD}} % non-italic subscript WD
\newcommand{\sdB}{_\textrm{sd}} % non-italic subscript sd
\newcommand{\comp}{_\textrm{comp}} % non-italic subscript comp

\title{A hot subdwarf--white dwarf super-Chandrasekhar candidate supernova Ia progenitor}
\author[1,2*]{Ingrid Pelisoli}
\author[3]{P. Neunteufel}
\author[1]{S. Geier}
\author[4,5]{T. Kupfer}
\author[6]{U. Heber}
\author[6]{A. Irrgang}
\author[6]{D. Schneider}
\author[1]{A. Bastian}
\author[7]{J. van Roestel}
\author[1]{V. Schaffenroth}
\author[8]{B.~N. Barlow}

\affil[1]{Institut f\"{u}r Physik und Astronomie, Universit\"{a}t Potsdam, Haus 28, Karl-Liebknecht-Str. 24/25, D-14476 Potsdam-Golm, Germany}
\affil[2]{Department of Physics, University of Warwick, Coventry, CV4 7AL, UK}
\affil[3]{Max Planck Institut f\"{u}r Astrophysik, Karl-Schwarzschild-Straße 1, 85748 Garching bei München}
\affil[4]{Kavli Institute for Theoretical Physics, University of California, Santa Barbara, CA 93106, USA}
\affil[5]{Texas Tech University, Department of Physics \& Astronomy, Box 41051, 79409, Lubbock, TX, USA}
\affil[6]{Dr. Karl Remeis-Observatory \& ECAP, Astronomical Institute, Friedrich-Alexander University Erlangen-Nuremberg (FAU), Sternwartstr. 7, 96049 Bamberg, Germany}
\affil[7]{Division of Physics, Mathematics and Astronomy, California Institute of Technology, Pasadena, CA 91125, USA}
\affil[8]{Department of Physics and Astronomy, High Point University, High Point, NC 27268, USA}

\affil[*]{ingrid.pelisoli@warwick.ac.uk}

\begin{abstract}
Supernova Ia are bright explosive events that can be used to estimate cosmological distances, allowing us to study the expansion of the Universe. They are understood to result from a thermonuclear detonation in a white dwarf that formed from the exhausted core of a star more massive than the Sun. However, the possible progenitor channels leading to an explosion are a long-standing debate, limiting the precision and accuracy of supernova Ia as distance indicators. Here we present HD 265435, a binary system with an orbital period of less than a hundred minutes, consisting of a white dwarf and a hot subdwarf --- a stripped core-helium burning star. The total mass of the system is $1.65\pm0.25$ solar-masses, exceeding the Chandrasekhar limit (the maximum mass of a stable white dwarf). The system will merge due to gravitational wave emission in 70 million years, likely triggering a supernova Ia event. We use this detection to place constraints on the contribution of hot subdwarf-white dwarf binaries to supernova Ia progenitors.
\end{abstract}

\begin{document}

\flushbottom
\maketitle

\thispagestyle{empty}

% Introduction + Results + Discussion limited to 3,000—3,500 words.
% 6-8 display items (figures and/or tables)
% Up to 50 references (excluding those cited exclusively in Methods)

% \section*{Introduction}
% Up to 500 words and no heading
Type Ia supernovae (SN~Ia) represent one of the crucial rungs on the cosmic distance ladder. As bright standard candles, they contribute to obtaining measurements of the Hubble constant $H_0$, which describes how fast the Universe is expanding at different distances \cite{schmidt1998, riess1998, perlmutter1999}. An accurate determination of the systematic uncertainties involved in these cosmological measurements requires a reliable identification of the progenitor channels contributing to the observed SN~Ia population. Current measurements of $H_0$ in the local Universe relying on SN~Ia \cite{riess2019} are inconsistent with estimates using the cosmic microwave background radiation observed by the Planck experiment \cite{planck2018}. In order to establish whether this $H_0$ tension \cite{bernal2016} is evidence for new Physics, or rather a consequence of poor determination of systematic uncertainties, it is imperative to understand possible SN~Ia channels and their relative contributions

Although the origin of SN~Ia has for long been understood as a thermonuclear detonation in a white dwarf \cite{hoyle1960}, triggered when a critical mass near the Chandrasekhar limit of 1.4~$M_{\odot}$ is reached, the mechanism for the explosion itself remains under debate \cite{hillebrandt2013}. Possible channels for achieving critical mass can be generally grouped in two: double degenerate, or single degenerate. In the double degenerate channel, the white dwarf has another compact star as a companion, and the detonation is triggered by the merger of the two objects \cite{whelan1973,iben1984,liu2018}. In the single degenerate channel, the white dwarf accretes mass from a companion up to a point in which thermonuclear explosion is triggered \cite{whelan1973,han2004}. Confirmed progenitors are extremely scarce for both channels \cite{rebassa2019}, making it challenging to explain observed rates \cite{maoz2012}. The once promising Henize 2-428 system \cite{santander2015} has recently been shown to have a total mass significantly lower than previously derived, and can no longer be considered as a SN~Ia progenitor \cite{reindl2020}. Even in the dedicated ESO supernovae type Ia progenitor survey (SPY), only two systems (WD2020-425 and HE2209-1444) have been identified as possible progenitors, both with sub-Chandrasekhar total masses \cite{napiwotzki2020}. The only known super-Chandrasekhar candidate progenitor is KPD~1930+2752 \cite{maxted2000}, a hot subdwarf with a close white dwarf companion. Another similar albeit less massive binary, CD-30$^{\circ}$11223 \cite{vennes2012, geier2013}, also qualifies as SN~Ia progenitor. These merging massive systems can also be of interest as gravitational wave sources, in particular as verification sources for the upcoming Laser Interferometer Space Antenna ({\it LISA}).

Here we report the discovery that HD~265435 is a candidate supernova progenitor and {\it LISA} verification binary composed by a hot subdwarf with a massive white dwarf companion. This $V=11.78$ binary system is at a distance of less than 500~pc from the Sun, making it the closest super-Chandrasekhar candidate supernova progenitor. We analysed the light curve obtained by the Transiting Exoplanet Survey Satellite (TESS)\cite{ricker2015} together with time-series spectroscopy to characterise the system and determine the component masses. The properties of this binary make it a candidate for both the single degenerate and double degenerate SN~Ia channels.

\section*{Results}

HD~265435 (TIC~68495594) was observed by TESS in Sector 20. The data revealed strong ellipsoidal variation, suggesting that the visible component of the system is tidally deformed by a compact object. The light curve also reveals pulsation frequencies showing rotational splitting. Following this discovery, we obtained time-series spectroscopy at the Palomar 200-inch telescope with the Double-Beam Spectrograph (DBSP)\cite{oke1982} covering one orbital cycle, with the aim of obtaining the radial velocity curve of the visible star. We also obtained high-resolution spectra with the Echellette Spectrograph and Imager (ESI) at the Keck II telescope to determine the line-of-sight rotational velocity, $v \sin i$. Combining the spectra with the radius estimate from fitting the spectral energy distribution (SED) using the {\it Gaia} early data release 3 (EDR3)\cite{gaia_edr3} parallax, we completely characterised the visible component, a hot subdwarf of spectral type OB (sdOB). The orbital inclination of the system was obtained from the estimated $v \sin i$, given the evidence that the hot subdwarf is tidally locked, and that in turn allowed us to constrain the mass of the unseen companion, which is likely a white dwarf with a carbon-oxygen core. We also fitted the TESS light curve without relying on any stellar parameters derived from the spectroscopy, obtaining a consistent solution (within 2-$\sigma$) that confirms the nature of the companion. The obtained stellar and binary parameters for HD~265435 are provided in Table~\ref{table:binary}.

\begin{table}[h!]
\caption{{\bf Astrometric, stellar, and orbital parameters for HD~265435.} Astrometric parameters are from {\it Gaia} EDR3\cite{gaia_edr3}, with a zero-point correction applied to the parallax\cite{edr3_zeropoint}. For stellar and orbital parameters, we show both the values obtained from spectroscopic analyses and from the light curve fit, if derived independently, as well as the adopted values which result from combining the two solutions by concatenating the distributions obtained for each parameter. Quoted values are the median, and uncertainties give the 68\% confidence interval. Where a $\chi^2$ fit was employed (see text), the systematic uncertainty was quadratically added to the statistical fit uncertainty. Quantities shown are the right ascension RA, declination DEC, parallax $\varpi$, proper motions in the right ascension, $\mu_\alpha$, and declination, $\mu_\delta$, directions, the hot subwarf effective temperature $T\eff\,{}\sdB$, surface gravity $\log~g\sdB$, helium abundance $\log~y\sdB$, mass $M\sdB$, radius $R\sdB$, and luminosity $L\sdB$, the white dwarf mass $M\WD$,
the zero-point of the ephemeris $T_0$, the orbital period $P$,  the radial-velocity semi-amplitude of the hot subdwarf $K\sdB$, the systemic velocity of the binary $V_0$, the mass ratio $q$, the orbital inclination $i$, the orbital distance $a$, the merging time due to gravitational waves $\tau$, and the gravitational wave amplitude $\mathcal{A}$.}
\label{table:binary}
\centering          
\begin{tabular}{c c c c} 
\hline\hline       
Parameter & Spectroscopic solution & Light curve solution & Adopted value\\
\hline
   RA (J2000) & \multicolumn{2}{c}{-} & 06:53:24.30 \\
   DEC (J2000) & \multicolumn{2}{c}{-} & +33:03:34.2 \\
%   $V$ & \multicolumn{2}{c}{-} & $11.78\pm0.18$\cite{hog2000} \\
%   $G$ & \multicolumn{2}{c}{-} & $12.0880\pm0.0021$\cite{gaia_edr3} \\
   $\varpi$ (mas) & \multicolumn{2}{c}{-} & $2.216\pm0.055$ \\
   $\mu_{\alpha}$ (mas/yr) & \multicolumn{2}{c}{-} & $-4.83\pm0.06$ \\
   $\mu_{\delta}$ (mas/yr) & \multicolumn{2}{c}{-} & $-4.583\pm0.0492$ \\
\hline
   $T\eff\,{}\sdB$ (K) & $34300\pm400$ & fixed & $34300\pm400$ \\
   $\log~g\sdB$ [cgs] & $5.62\pm0.10$ & $5.52\pm0.04$ & $5.55^{+0.12}_{-0.06}$  \\
   $\log~y\sdB$ & $-1.46\pm0.10$ & \multicolumn{2}{c}{-} \\
   $M\sdB$ ($M\sun$) & $0.62^{+0.17}_{-0.13}$ & $0.64^{+0.10}_{-0.09}$ & $0.63^{+0.13}_{-0.12}$ \\  
   $R\sdB$ ($R\sun$) & $0.203\pm0.006$ & $0.230\pm0.012$ & $0.213^{+0.025}_{-0.013}$ \\
   $L\sdB$ ($L\sun$) & $51\pm4$ & $67^{+8}_{-7}$ & $57^{+14}_{-8}$ \\
   $M\WD$ ($M\sun$) & $0.91^{+0.11}_{-0.10}$ & $1.10\pm0.11$ & $1.01\pm0.15$ \\  
\hline
   $T_0$ (BJD)   & fixed & \multicolumn{2}{c}{$24571909.6899552(26)$} \\
   $P$ (days)    & fixed & \multicolumn{2}{c}{$0.0688184888(32)$} \\
   $K\sdB$ (km/s) & $343.1\pm1.2$ & Gaussian prior & $343.1\pm1.2$  \\
   $V_0$ (km/s)  & $8.2\pm0.8$ & \multicolumn{2}{c}{-} \\
   $q$           & $1.46^{+0.22}_{-0.18}$ & $1.70^{+0.11}_{-0.09}$ & $1.63^{+0.15}_{-0.26}$ \\
   $i$ (degrees) & $76\pm6$ & $60\pm2$ & $64^{+14}_{-5}$ \\
   $a$ ($R\sun$) & $0.805^{+0.048}_{-0.042}$ & $0.851\pm0.034$ & $0.831^{+0.043}_{-0.050}$ \\
   $\tau$ (Myr) & $76^{+25}_{-19}$ & $63^{+15}_{-11}$ & $70^{+26}_{-16}$ \\
   $\mathcal{A}$ ($10^{-22}$) & $3.1^{+1.1}_{-0.8}$ & $3.7^{+0.8}_{-0.7}$ & $3.5^{+1.0}_{-0.9}$ \\
   \hline                  
\end{tabular}
\end{table}

\subsection*{Period determination}

\begin{figure}[h!]
\centering
\includegraphics[width=0.75\hsize]{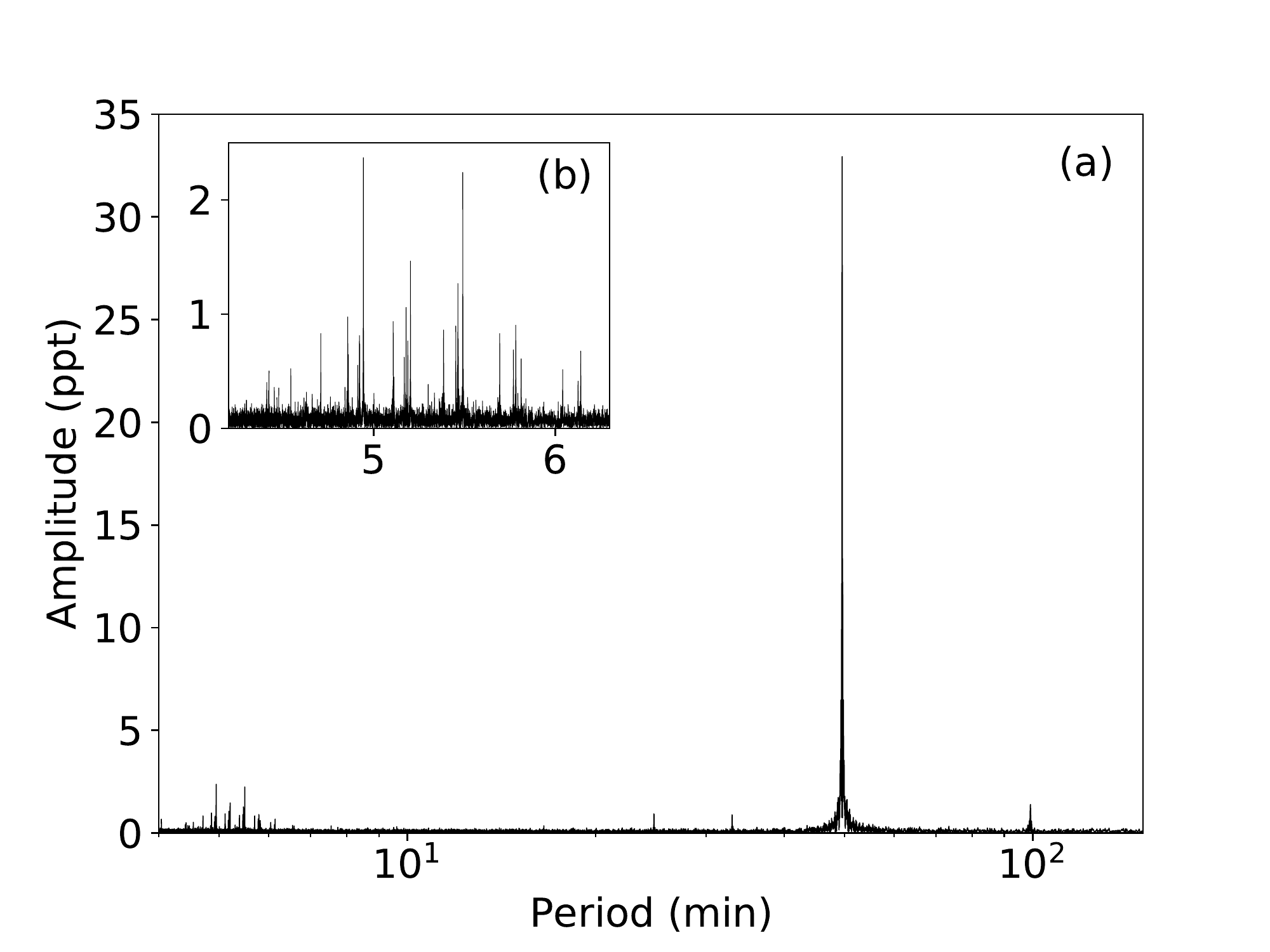}
\caption{{\bf Periodogram of the TESS light curve.} Panel (a) shows the full periodogram, in which a dominant peak can be seen at 49.54959(15)~min, corresponding to half the orbital period. Multiple low-period peaks can also be identified in the range of 4--6 minutes, as detailed in the inset panel (b).}
\label{periodogram}
\end{figure}

A Lomb-Scargle periodogram of the TESS light curve showed a dominant peak at 49.54959(15)~min (Fig.~\ref{periodogram}). Phase-folding the light curve to twice this dominant peak revealed the occurrence of Doppler boosting \cite{shakura1987}, causing a height difference of $\approx 0.3$\% between consecutive maxima, indicating that the real orbital period is $P = 99.09918(29)$ minutes. A smaller amplitude peak can be seen at this period, as well as harmonics at P/3 and P/4. The periodogram also shows a wealth of short-period peaks, which are in the correct range for $p$-mode pulsations of the hot subdwarf \cite{charpinet1996, kilkenny1997}. We identified a total of 33 frequencies above a detection level of five times the average amplitude of the Fourier transform (see Supplementary Table~\ref{table:pmods}). Many of these frequencies are part of rotational multiplets, which result from the spherical symmetry being broken by rotation\cite{kawaler2005}. The separation is expected to be proportional to the rotation period, and close to equal to it for $p$-mode oscillations \cite{reed2014}. We find the separation between peaks to be equal to the orbital period (see Supplementary Figure~\ref{pmodes}), suggesting thus that the rotation period of the subdwarf is equal to the orbital period, i.e. the system is synchronised, as is often is observationally inferred for hot subdwarf binaries with orbital periods of less than half a day \cite{geier2008}.

\begin{figure*}[h!]
\centering
\includegraphics[width=\hsize]{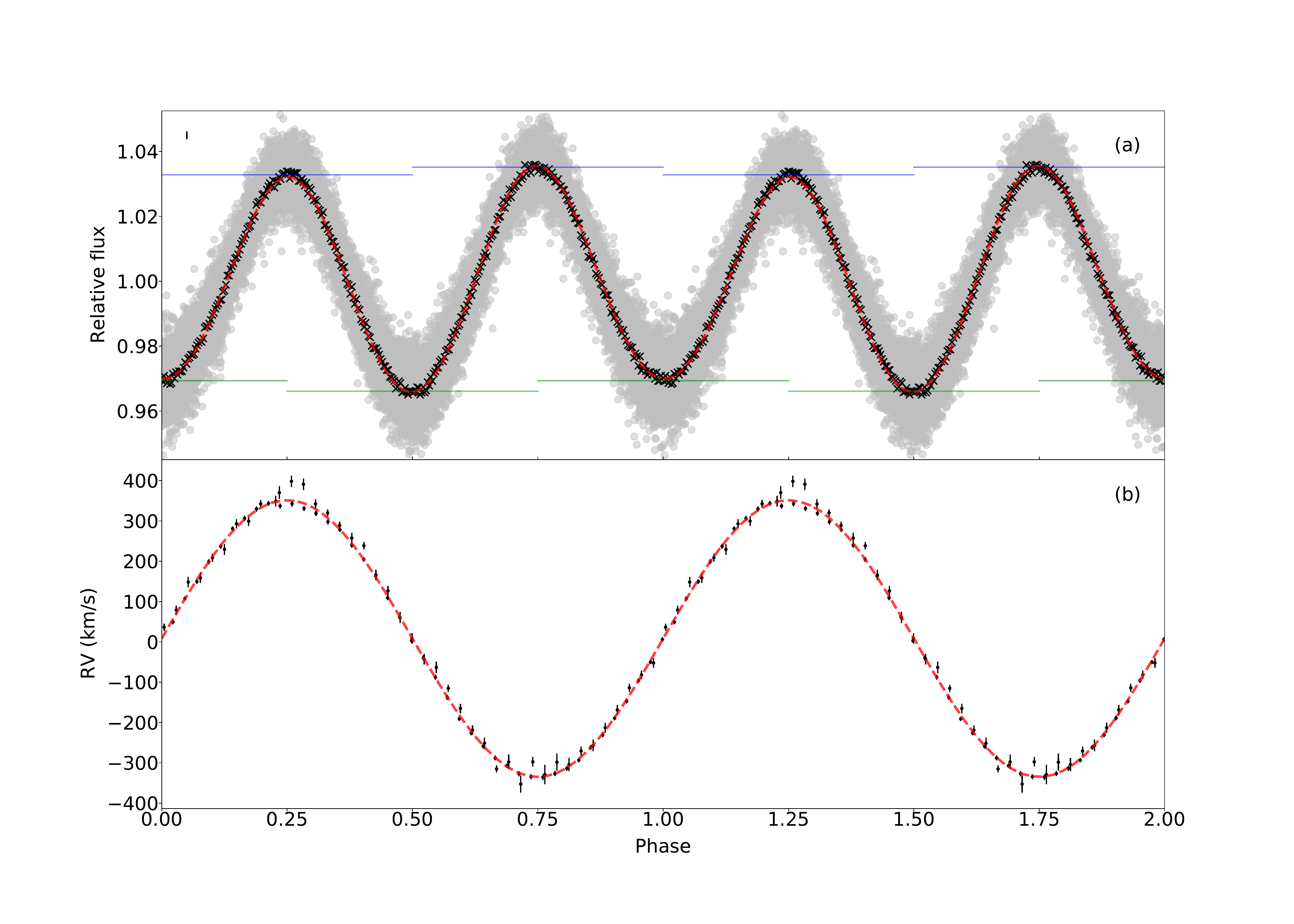}
\caption{{\bf Phased data for HD~265435.} Panel (a) shows the light curve, and panel (b) the radial velocity data, both phase-folded to the obtained ephemeris. In panel (a), the original TESS data is shown as grey dots. The phased data were binned every 50 points for the light curve fit, as shown by the black crosses. The median one-sigma uncertainty for the binned data is shown as a bar to the top left. In panel (b), the individual radial velocity estimates are shown in black with their one-sigma uncertainties. The best-fit model is shown as red dashed lines in both panels. The dominant effect in the light curve is an ellipsoidal variation caused by the tidal deformation of the visible component. The tidal deformation also leads to gravity darkening, causing the two minima to show different depths. The less prominent effect of Doppler boosting, which causes maxima to show different heights, is in turn caused by the orbital motion of the visible companion. We trace the location of each minima and maxima with solid horizontal lines, to make the effect of these phenomena more easily noticeable.}
\label{twophase}
\end{figure*}

The TESS 2-minute cadence is not adequate for correctly sampling such short periods, therefore we attempt no asteroseismologial analysis. These periods were identified so that their effect could be subtracted from the light curve prior to modelling the effect of the binary companion, otherwise they would have lead to systematic errors on the final fit parameters. This was done recursively: we first calculated a preliminary model for the variability due to binarity (see Methods for details), which we then subtracted from the original light curve in order to determine the short periods. Next we performed a global fit using all 33 identified peaks, and subtracted the obtained model from the original light curve. The preliminary model was also used to fit the full light curve in order to refine the period and determine the zero point of the ephemeris (adopted here as the superior conjunction of the unseen companion). We obtained a period of $P = 0.0688184888(32)$~days, and $BJD_0 = 24571909.6899552(26)$~days. The light curve and radial velocity data folded using this ephemeris is shown in Fig.~\ref{twophase}.

\subsection*{The radial-velocity curve of the hot subdwarf}

We determine the radial velocities by cross-correlating each spectrum obtained with DBSP with a best-fit spectral template (see the Methods section for a full description of the procedure). We analysed spectra from the blue and red arms separately, as they are not obtained simultaneously, and obtained consistent radial velocities. We fitted the radial velocities assuming a circular orbit, with the period fixed to the photometric period, as the time span of our radial velocity curve would not allow a precise independent determination of the period. We allowed the zero point of the ephemeris to vary by $P/2$ in order to account for possible phase shifts between the photometric and spectroscopic data, obtaining a value consistent with the photometry within four decimal places. The best-fit model is shown in the bottom panel of Fig.~\ref{twophase}. The obtained radial velocities revealed a large radial semi-amplitude of $K\sdB = 343.1\pm1.2$~km~s$^{-1}$, implying a mass function for the unseen companion of
\begin{eqnarray}
f\comp = \frac{M\comp^3 \sin^3(i)}{(M\sdB + M\comp)^2} = \frac{P\,K\sdB^3}{2\pi G} = 0.288\pm0.003~M\sun.
\label{bin_func}
\end{eqnarray}
Combining the obtained systemic velocity with the {\it Gaia} astrometric information, we find the system to show dynamics consistent with the thin disk of the Galaxy (see Methods for details).

\subsection*{Characterising the system and the nature of the companion}

The obtained spectra revealed the visible component to be a sdOB, as already suggested based on its {\it Gaia} DR2 parameters \cite{geier2019}. We performed spectral fits of the Doppler corrected DBSP spectra (as detailed in the Methods section), obtaining an effective temperature of $T\eff = 34300\pm400$~K and surface gravity with $\log~g = 5.62\pm0.10$.  With $T\eff$ and $\log~g$ fixed, we obtained $v \sin i$ and helium-to-hydrogen ratio (by number) $\log~y$ by fitting each of the high-resolution ESI spectra separately, to avoid additional broadening introduced by co-adding the spectra. This resulted on $\log~y = -1.46\pm0.04$ and $v \sin i = 152\pm6$~km\,s$^{-1}$.

Performing a fit to the SED (see Methods for further details), we find the photometry to be consistent with a single hot subdwarf, finding no contribution from the unseen companion. Using the {\it Gaia} EDR3 parallax, our SED fit provided a radius estimate of $R\sdB = 0.203\pm0.006~R\sun$, implying a hot subdwarf mass of $0.62^{+0.17}_{-0.13}~M\sun$ from the obtained $\log~g$. Given the indication that the rotational period of the hot subdwarf is synchronised with the orbital period, the orbital inclination can also be obtained from radius and $v \sin i$, which give $76\pm6^{\circ}$. Finally, Eq.~\ref{bin_func} given above can be solved to obtain the mass of the unseen companion, which is found to be $0.91^{+0.11}_{-0.10}~M\sun$, corroborating its nature as a compact object, as no contribution from an early-type companion is observed.

Alternatively, the multiple effects observed in the light curve of HD~265435 can be used to constrain some stellar parameters of the system independently from the spectroscopy. The ellipsoidal variation, gravity darkening and Doppler boosting effects depend mainly on the radius of the hot subdwarf and masses of both components. The temperature and radius of the unseen companion, on the other hand, can still not be constrained, as its contribution to the light curve is negligible and no eclipses are observed.

We fit the light curve using {\sc lcurve} \cite{copperwheat2010}, a code that uses a inhomogeneous grid of points, optimised to reproduce the stellar surface area, to model the brightness of two orbiting stars with shapes set by a Roche potential. We left as free parameters the mass ratio $q$, the inclination angle $i$, the scaled equatorial radius of the hot subdwarf $r\sdB = R\sdB/a$, where $a$ is the orbital distance, and the velocity scale, $V_{\textrm{scale}} = (K\sdB+ K\comp)/\sin~i$. The value of $K\sdB$ was required to be consistent with the determination from the radial velocity observations, but no other priors were applied (see Methods for details on the procedure).

\begin{figure*}[h!]
\centering
\includegraphics[width=\hsize]{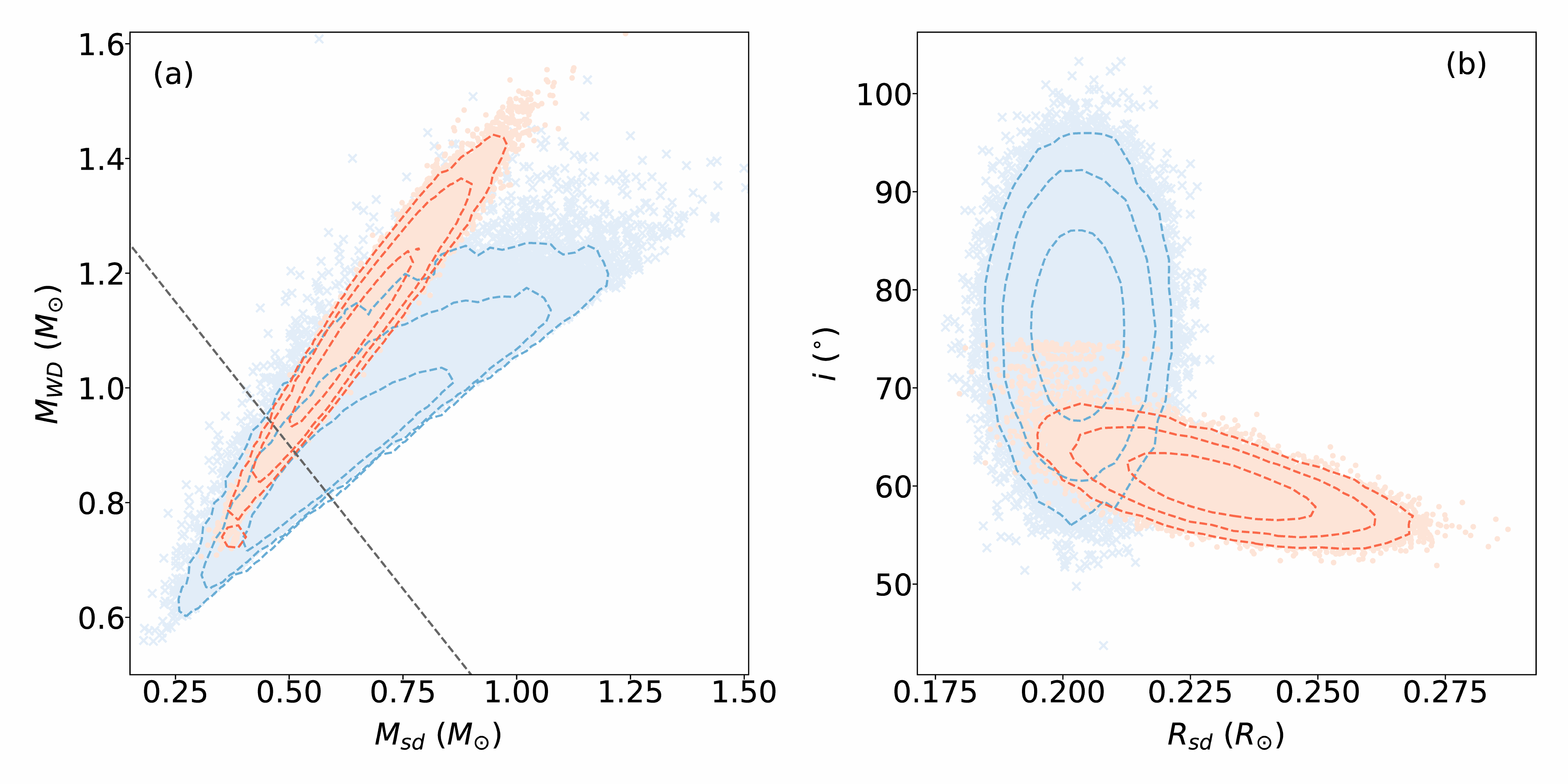}
\caption{{\bf Comparison between photometric and spectroscopic solutions.} Panel (a) shows the component masses, and panel (b) the hot subdwarf radius and the inclination of the system. The blue crosses show values derived by spectroscopic analyses, whereas the orange dots result from the light curve solution. The dashed contours show confidence intervals corresponding to 68\%, 95\% and 99\%. Mass combinations above the dashed grey line in panel (a) are above the Chandrasekhar limit.}
\label{solcomp}
\end{figure*}

We obtained $q = 1.70^{+0.11}_{-0.09}$, $i = 60\pm2^{\circ}$, $r\sdB = 0.289\pm0.009$, and $V_{\textrm{scale}} = 625\pm25$~km~s$^{-1}$. These parameters imply masses of $M\sdB = 0.64^{+0.10}_{-0.09}~M\sun$ and $M\comp = 1.10\pm0.11~M\sun$, and a hot subdwarf radius of $R\sdB = 0.232\pm0.012~R\sun$, therefore consistent with the parameters derived from spectroscopic fitting within their 95\% confidence intervals, as illustrated in Fig.~\ref{solcomp}.

Combining the two consistent solutions by concatenating the distributions obtained for each parameter, we find the stellar and orbital parameters given in the third column of Table~\ref{table:binary}. The mass of the companion is found to be $M\comp = 1.01\pm0.15~M\sun$. This implies that the companion is likely a white dwarf with a C/O core, although an O/Ne(/Mg) composition is possible if the mass is above 1.088$~M\sun$ \cite{lauffer2018}. The total mass of the system is found to be $1.65\pm0.25~M\sun$.

\subsection*{Future evolution of HD~265435}

The evolution of a binary is primarily determined by the total mass of the system, initial orbital separation and the evolutionary status of the hot subdwarf at the time of Roche-lobe overflow (RLOF). The obtained radius hints at a hydrogen envelope with a current mass around $1.5 \times 10^{-4}~M\sun$, which is typical for hot subdwarf stars \cite{heber2016}.

\begin{figure}[h!]
\centering
\includegraphics[width=0.95\hsize]{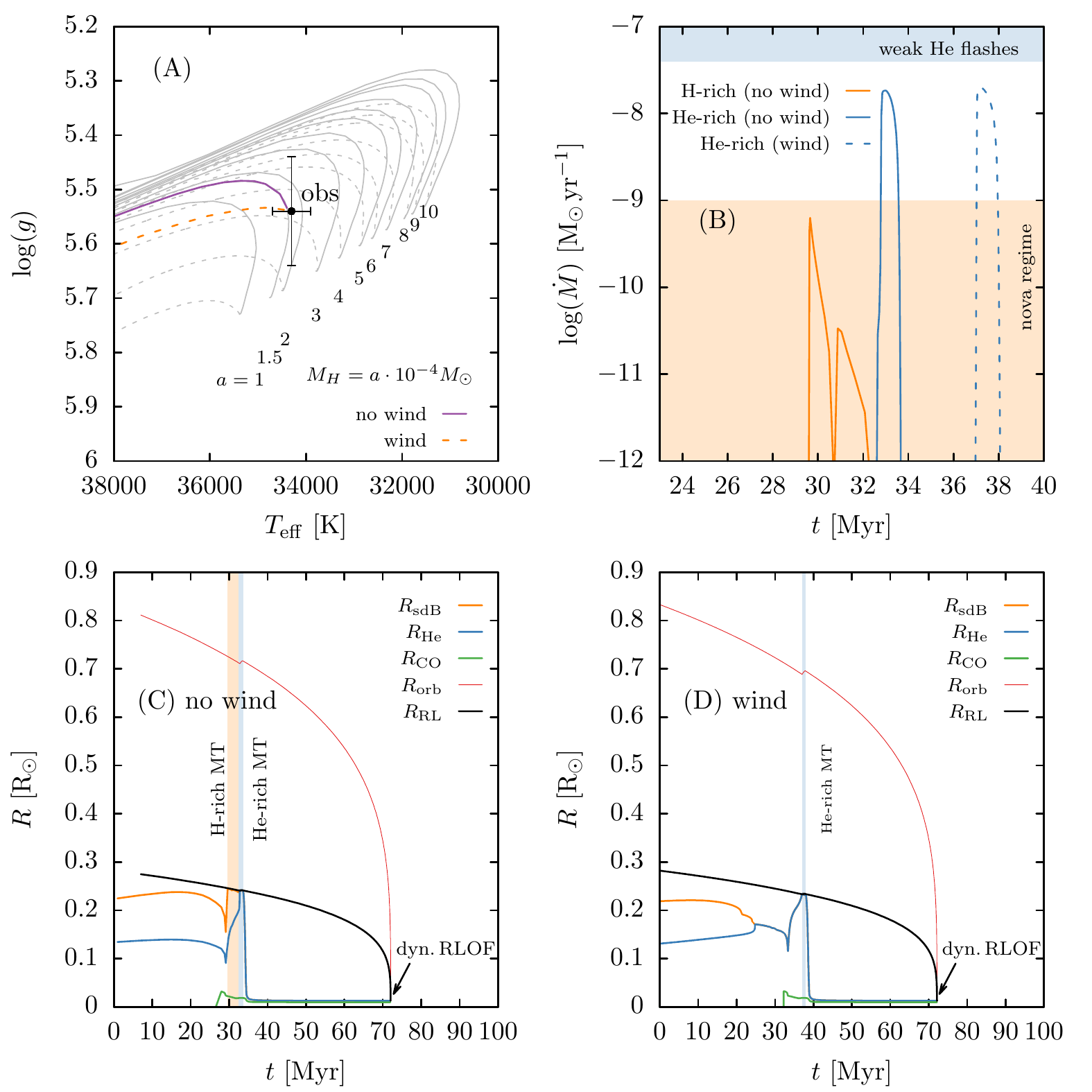}
\caption{{\bf Model and prediction of the future evolution of HD265435.} Panel (A) shows predicted evolutionary tracks in $T_\mathrm{eff}$-$\log(g)$ for hot subdwarfs of a mass of $0.63\,\text{M}_\odot$ and H-envelopes between $10^{-4}~M\sun$ and $10^{-3}~M\sun$ with the solid purple line representing the favoured models with $\sim 1.5 \times 10^{-4}~M\sun$ (no wind) starting from the beginning of the binary run. The weak wind model with an initial hydrogen envelope of $\sim 1.5 \times 10^{-4}~M\sun$ is represented by the dashed orange line. The hot subdwarf observed position in the diagram is as indicated, with error bars representing the systematic uncertainty of the obtained spectral parameters.
Panel (B) shows the evolution of the system's mass transfer rate under the assumption of either no wind (solid lines) and a weak wind (dashed). Colour indicates whether the transferred mass is H or He enriched. Shaded areas indicate where H-accretion induced novae or He-accretion induced flashes are expected.
Panel (C) shows the orbital evolution of the benchmark model system without wind. Here, $R_\text{sdB}$ is the total radius of the hot subdwarf, including hydrogen envelope. $R_\text{He}$ is the radius of the He-rich shell and $R_\text{CO}$ the radius of the inert CO core. $R_\text{RL}$ is the radius of the Roche lobe of the hot subdwarf. The shaded areas indicate expected mass transfer phases and dominant composition of the transferred material.
Panel (D) is equivalent to Panel (C) but for the weak wind model.}
\label{model}
\end{figure}

We carried out numerical simulations of the evolution of the system in order to determine its possible outcomes. Assuming solar metallicity, a helium star with a total mass of $0.63~M\sun$ and a remaining H-envelope of $10^{-4}~M\sun$ about halfway through its expected core He burning lifetime yields physical parameters consistent with the observed values (see panel (A) of Fig.~\ref{model}). The model was placed in a binary with a carbon-oxygen core white dwarf approximated by a point mass. We include the effects of rotation, assuming tidal locking and angular momentum loss through gravitational radiation. Our benchmark model assumes no wind mass loss, which is the standard assumption in modelling of hot subdwarfs. However, as our results are partially sensitive to the occurrence of winds, which are, in the case of hot subdwarf stars, still a matter of debate, we also include a weak wind \cite{Unglaub2008} in an alternative model. We note that inclusion of wind suggests an initial mass of the hydrogen envelope of $3 \times 10^{-4}~M\sun$, half of which has been ejected at the time of observation. Further details of the simulation are given in the Methods section.

We find that RLOF is precipitated by the end of the hot subdwarf's core helium burning phase after $\sim 29.6$~million years (see panels (B) and (C) of Fig.~\ref{model}). In our benchmark model, subsequent expansion then leads to RLOF. The transferred material will be hydrogen enriched for the first $\sim 3.0$~million years of RLOF, subsequently becoming He-enriched, as the remaining hydrogen envelope is stripped. The He-enriched phase is expected to last for $\sim 1$~million years, resulting on $\sim 0.015~M\sun$ of He-rich material being transferred. Mass transfer rates are expected to lie in the range $10^{-11}-10^{-9}~M\sun/\mathrm{yr}$, enough to indicate a phase of classical nova eruptions \cite{iben1992,shara2018}, and will not exceed $2.5 \times 10^{-8}~M\sun/\mathrm{yr}$ during the He-rich phase, which indicates that helium will be accumulated quiescently, without igniting \cite{woosley2011, neunteufel2016}, on the white dwarf.

The introduction of a weak wind has the effect of delaying the RLOF phase, which happens then after $\sim 37$~million years (see panels (B) and (D) of Fig.~\ref{model}). This discrepancy in the onset of RLOF is explained firstly by the benchmark model requiring $\sim 5$~million years longer to acquire the observed properties and, secondly, by the presence of a H-enriched envelope, which is removed by winds in the alternative scenario. This envelope expands faster than the H-depleted parts of the envelope as the star moves into He-shell burning. The expansion of the helium envelope preceding the end of core helium burning in the alternative model is a result of the removal of the hydrogen rich envelope by the weak wind, to which the helium envelope, in preserving the star's boundary conditions and smooth pressure gradient, reacts by expanding. We note that our qualitative and quantitative predictions for the future evolution of this system are otherwise unaffected by the presence of a wind.

This mass transfer rate is sufficient to stabilise the binary against further inspiral due to gravitational wave radiation (GWR) for the duration of the mass transfer phase, leading to an increase of the merger time ($71.8$~Myr according to our simulation) \cite{Tutukov1979,neunteufel2020}. Quiescent accumulation is a prerequisite for ignition of a thermonuclear SN according to the double detonation mechanism (for more detail see Methods section), however, the amount of transferred material is too small.
The end of this mass transfer phase is precipitated by the remaining helium envelope of the hot subdwarf losing sufficient mass, both due to nuclear burning and mass transfer to the companion, for further helium burning to become unsustainable. At this point the hot subdwarf will contract thermally to become a CO white dwarf with a remnant He envelope of $\sim 0.03~M\sun$, i.e. a hybrid HeCO white dwarf.

\begin{figure}[h!]
\centering
\includegraphics[width=0.75\hsize]{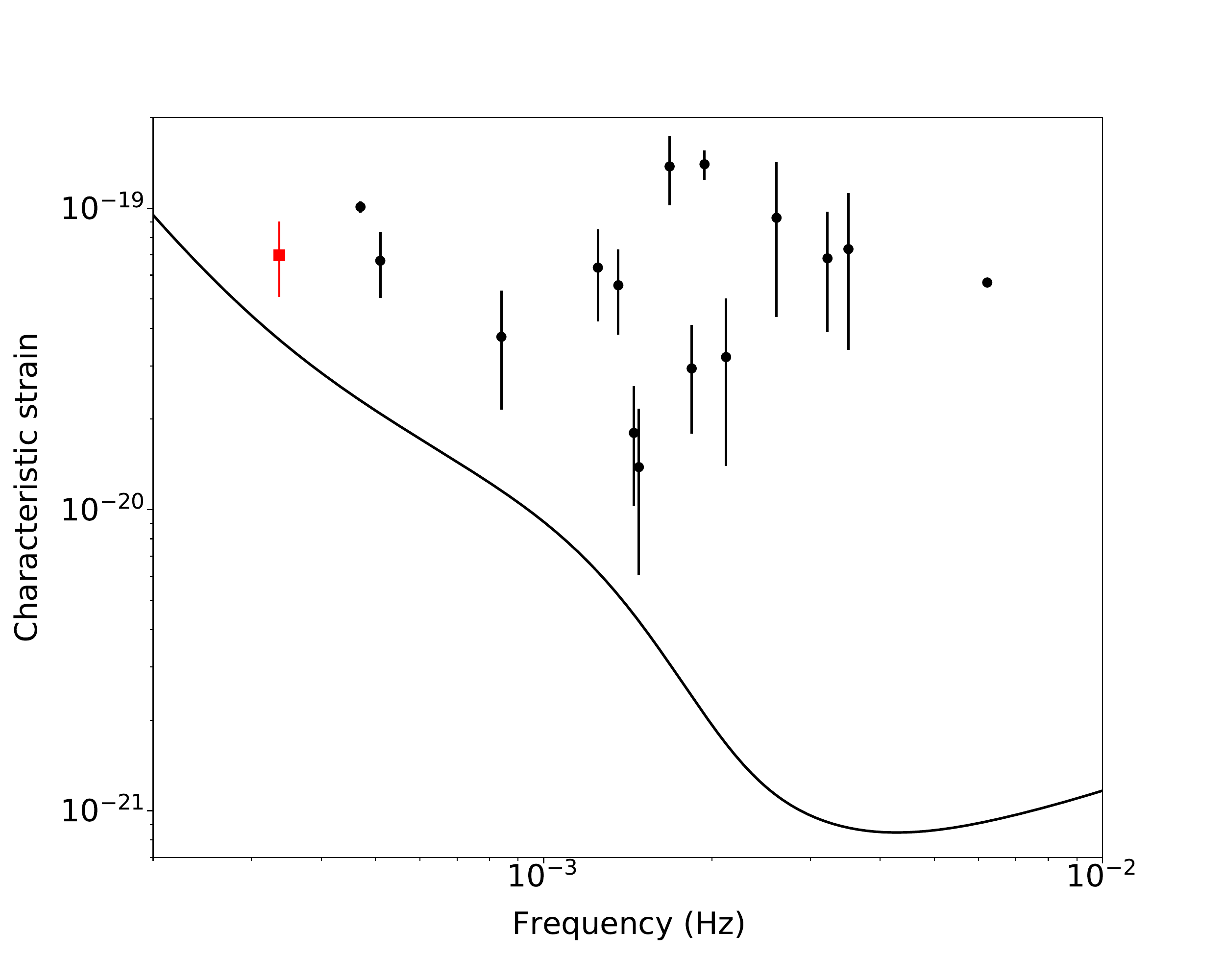}
\caption{{\bf Gravitational wave frequency and strain of HD~265435.} The black line shows the {\it LISA} sensitivity curve for a four-year mission \cite{robson2019}. The red square shows the median strain and frequency of HD~265435, with the errorbar representing the 68\% confidence interval. Black data points are previously known verification binaries \cite{kupfer2018}.}
\label{strain}
\end{figure}

Following this mass transfer phase, the system will continue to lose angular momentum due to GWR. The former hot subdwarf will then fill its Roche lobe once again. With a mass ratio of $q \approx 1.64$, this episode will likely lead to dynamically unstable RLOF, in the course of which the former hot subdwarf is disrupted and merges with the heavier companion, resulting on one of three possible channels for the thermonuclear detonation, (i) a prompt detonation \cite{Kromer2010} of the more massive white dwarf, (ii) a violent merger of the two white dwarfs \cite{pakmor2010}, or (iii) unstable ignition of helium on the more massive white dwarf travelling along the accretion stream and leading to the double detonation of the white dwarf donating mass \cite{pakmor2021}. However, we emphasise the presence of a non-negligible amount of unburnt helium on the accretor. The derived masses allow us to constrain the merger time due to gravitational wave emission \cite{kraft1962}, which is found to be $70^{+26}_{-16}$~Myr, consistent with our numerical simulation. The characteristic strain\cite{shah2012, moore2015} of the system places it above the detection limit of {\it LISA}, as illustrated in Fig.~\ref{strain}.

We note that, given our obtained mass intervals, there is a $\simeq$ 16\% probability that the total mass of the system is below the Chandrasekhar mass. In this case, the system would not lead to a supernovae through any standard double degenerate or single degenerate channel. The likely scenario is recurrent novae starting some 20 Myr from the time of observation, followed by a double degenerate dynamical merger. There remains a possibility for prompt-detonation, which depends on the presence of helium on the former hot subdwarf \cite{Kromer2010,gronow2020}.

\section*{Discussion}
% One or two short paragraphs

The newly discovered system HD~265435 brings the number of hot subdwarf with white dwarf companions that qualify as supernova progenitors to three, making this the class of binaries with the most observed progenitor candidates. These supernova may ultimately not show typical SN~Ia spectra, but may appear subluminous and/or peculiar depending on their mass ratio \cite{pakmor2010}. HD~265435 has very similar properties to KPD~1930+2752: both harbour a relatively hot subdwarf star ($T\eff > 30\,000$~K) and a massive white dwarf with a CO core as companion, bringing the total mass of the system above the Chandrasekhar limit. Additionally, both hot subdwarfs in HD~265435 and in KPD~1930+2752 have been observed to show peaks in the range of $p$-mode pulsations. CD-30$^{\circ}$11223, on the other hand, has lower mass components and total mass slightly below the Chandrasekhar limit. However, KPD~1930+2752 will likely evolve through the core-He burning phase without filling its Roche lobe and transferring mass to the companion\cite{neunteufel2016}, whereas mass transfer is predicted to happen to both HD~265435 and CD-30$^{\circ}$11223. Therefore, in terms of its evolutionary fate, HD~265435 is more similar to CD-30$^{\circ}$11223. A class of Roche lobe-filling hot subdwarf binaries has recently been discovered \cite{kupfer2020b}, providing observational evidence for the existence of systems undergoing mass transfer before the hot subdwarf evolves into a white dwarf.

Perhaps the most remarkable common property of these candidate supernova progenitors is the fact that they are all found within 1~kpc of the Sun and seem to be members of the thin disk, showing relatively low Galactic latitudes. Given their {\it Gaia} EDR3 zero-point corrected parallaxes, CD-30$^{\circ}$11223 is at $349\pm6$~pc and $b = 28.9^{\circ}$, HD~265435 at $451\pm11$~pc and $b = 14.8^{\circ}$, and KPD~1930+2752 at $825^{+27}_{-24}$~pc and $b = 4.3^{\circ}$. Making the assumption that these three objects consist of the entire sample of hot subdwarf-white dwarf binaries that qualify as supernova progenitors within 1~kpc and taking into account a Poissonic uncertainty, that would imply a space density of $0.22\pm0.13$~kpc$^{-3}$ for this type of system, considering the effective volume given by the thin disk density \cite{juric2008}. We can also roughly estimate the rate of SN~Ia that can be attributed to such systems. There are $\sim 3000$ hot subdwarf candidates within 1~kpc \cite{geier2019}. Accounting for an estimated contamination level of 10\%\cite{geier2019}, this would suggest that 3 out of the 2700 hot subdwarfs within 1~kpc are possible SN~Ia progenitors. Given the birthrate of such stars of 0.014--0.063~yr$^{-1}$ \cite{han2003}, this implies that the SN~Ia rate that can be attributed to hot subdwarf-white dwarf binaries is 1.5--7$\times 10^{-5}$~yr$^{-1}$. Population synthesis simulations suggest a larger value of $\sim 3 \times 10^{-4}$~yr$^{-1}$ for the contribution of helium star-white dwarf binaries to the SN~Ia rate \cite{wang2010}, but this estimate includes also helium stars more massive than hot subdwarfs. Our estimate is comparable to the estimated contribution from double degenerate white dwarf binaries, which is $2.1\pm1.0\times 10^{-5}$~yr$^{-1}$ \cite{rebassa2019}. The Galactic SN~Ia rate is in turn estimated to be $7.2\pm2.3\times 10^{-3}$~yr$^{-1}$ \cite{li2011, maoz2014}. Therefore, our estimate suggests that hot subdwarf-white dwarf binaries cannot bring the Galactic SN~Ia rate into agreement with observed progenitor rates, despite being the most numerous observed class of progenitors.  Our estimate should be, however, regarded as a lower limit, since we have assumed that there are no other SN~Ia progenitors consisting of hot subdwarf-white dwarf binaries within 1~kpc. The TESS extended mission, as well as future missions such as the Legacy Survey of Space and Time (LSST), will put this assumption to test.

%TC:ignore

\section*{Methods}
% Ideally not exceed 3,000 words but may be longer if necessary

\subsection*{Observations and data reduction}

HD~265435 (TIC~68495594) was observed by TESS in Sector 20, yielding two-minute cadence data over a baseline of 26.3~days, with a three day gap after 12.3~days during which the data were being downloaded to Earth. We retrieved the light curve derived by the TESS Science Processing Operations Center (SPOC), and used the PDCSAP flux, which corrects the simple aperture photometry (SAP) to remove instrumental trends and contributions to the aperture expected to come from neighbouring stars identified in a pre-search data conditioning (PDC). The pipeline also provides an estimate of the contribution of the target to the flux in the aperture, taking into account possible contamination by neighbouring targets. The value for HD~265435 is 0.65. The contamination is likely due to a $G = 12.3$ star 28" away, given the TESS pixel size of 21".. This much redder star ($G_{BP} - G_{RP} = 0.786$, compared to $G_{BP} - G_{RP} = -0.469$ for HD~265435) is likely a main sequence F star given the stellar parameters in the {\it Gaia} DR2 ($T\eff = 5860$~K and $R = 1.50~R\sun$\cite{andrae2018}). We identify no variability that could be attributed to this contaminating source, therefore we assume it to be constant, and that light curve amplitude has been correctly corrected by the SPOC pipeline.

Optical spectra were obtained at the Palomar 200-inch telescope with DBSP using a low resolution mode ($R \simeq 1200$). We obtained 40 exposures of 120~seconds covering 1.65~hours on March 02 2020. An average bias and normalised flat-field frame was made out of 10 individual bias and 10 individual lamp flat-fields. To account for telescope flexure, an arc lamp was taken at the position of the target after each observing sequence. For the blue arm, FeAr arc exposures were taken, and HeNeAr for the red arm. Both arms of the spectrograph were reduced using a custom \texttt{PyRAF}-based pipeline \cite{bellm2016}. The pipeline performs standard image processing and spectral reduction procedures, including bias subtraction, flat-field correction, wavelength calibration, optimal spectral extraction, and flux calibration. The trailed spectra are shown in Supplementary Fig.~\ref{trail_spec}, and clearly show periodical changes of the line centres. 
Finally, we obtained ten 60~second exposures with ESI at the Keck II telescope on September 10 2020, which were combined into a $R \simeq 6000$, high signal-to-noise ratio ($S/N \simeq 170$) spectrum. ThAr arc exposures were taken at the end of the night. The spectra were reduced using the \texttt{MAKEE} pipeline following the standard procedure: bias subtraction, flat fielding, sky subtraction, order extraction, and wavelength calibration.

\subsection*{The spectral fit of the hot subdwarf}

The observed spectra were matched to a model grid by $\chi^2$ minimisation \cite{2014A&A...565A..63I}. The quantitative spectral analysis is based on a new grid of model atmospheres and synthetic hydrogen and helium spectra that account for deviations from local thermodynamic equilibrium (LTE). We start from an LTE temperature/density stratification calculated with the {\sc Atlas12} code \cite{1996ASPC..108..160K}. Non-LTE population numbers of hydrogen and helium levels are then calculated with the {\sc Detail} code \cite{1981PhDT.......113G,detailsurface2} and handed back to {\sc Atlas12} to correct the atmospheric structure for non-LTE effects in an iterative process \cite{2018A&A...615L...5I}. After convergence, {\sc Detail} is used again to solve the coupled equations of radiative transfer and statistical equilibrium numerically. Finally the {\sc Surface} code \cite{1981PhDT.......113G,detailsurface2} is used to calculate the emergent spectrum using the non-LTE occupation numbers and detailed line-broadening tables. This hybrid approach has been shown to reproduce observations of B-type stars \cite{przybilla2011}, and has since been has applied to the entire $T\eff$ and $\log~g$ range of sdB and sdOB stars, from the coolest ($\approx$ 23000\,K)
\cite{silvotti2019, sahoo2020} to the very hottest ($\approx$ 40000\,K) \cite{silvotti2021}. Recent updates to all three codes \cite{2018A&A...615L...5I} such as the implementation of the occupation probability formalism \cite{1994A&A...282..151H} for hydrogen, and Stark broadening tables for hydrogen and neutral helium \cite{2009ApJ...696.1755T,1997ApJS..108..559B} are considered. We first fitted the individual DBSP spectra, which have better normalisation and cover the Balmer jump, to determine $T\eff$ and $\log~g$; $v \sin i$ was also left as a free parameter. We find $T\eff$ = $34300\pm400$~K and $\log g$ = $5.62\pm0.10$. Our $\log~g$ estimate differs from previous literature results\cite{lei2018}, because of orbital smearing of their spectra, which were exposed for $\approx$ 1/3 of the orbital period. Next we fit each of the higher-resolution ESI spectra to determine $v \sin i$ and $\log~y$, fixing the values of $T\eff$ and $\log g$ to those determined from the DBSP spectra. We fit only the helium lines, which are more sensitive than the Balmer lines, obtaining $\log~y$ = $-1.46\pm0.10$ and $v \sin i  = 152\pm7$~km~s$^{-1}$. Exemplary spectroscopic fits are shown in Supplementary Fig.~\ref{spec_fit}.

\subsection*{The SED fitting method}

The observed magnitudes were matched to a synthetic flux distribution employing a $\chi^2$ based fitting routine \cite{2018OAst...27...35H}. The synthetic flux distribution is interpolated from a grid of model SEDs calculated with {\sc Atlas12}, as described in the previous section, for the spectroscopically inferred atmospheric parameters. With the SED fit, we derive the angular diameter $\Theta = 2R/d$ (with the stellar radius $R$ and distance $d$), a scaling factor derived from the observed flux $f(\lambda)$ and the synthetic stellar surface flux $F(\lambda)$ by making use of the geometric flux dilution, that is, $f(\lambda)=\Theta^2 F(\lambda)/4$. The radius can then be determined from $R = \Theta/(2\varpi)$ using the trigonometric parallax measurement $\varpi$ provided by the {\it Gaia} EDR3, which has high-precision and quality indicators within specifications, in particular the re-normalised unit weight error \cite{RUWE} (RUWE).

As HD~265435 is located at low Galactic latitude ($b = +14.8^\circ$), interstellar reddening must be taken into account by fitting the interstellar colour excess along with the angular diameter. To model interstellar absorption we use the reddening law of Fitzpatrick et al. \cite{2019ApJ...886..108F}. We find interstellar reddening to be relatively small, with E(B-V)= 0.044 $\pm$ 0.011, compared to values from reddening maps \cite{1998ApJ...500..525S, 2011ApJ...737..103S}, implying that most of the line-of-sight reddening occurs beyond HD~265435. We derive the stellar radius from the angular diameter $\Theta$ and the parallax $\varpi$ by accounting for a {\it Gaia} EDR3 zero point offset of $-0.049$~mas, calculated for HD265435\cite{edr3_zeropoint}, and find $R = 0.203\pm0.06~R\sun$. Our fit is shown in Supplementary Fig.~\ref{fig:photometry_sed_single}.

\subsection*{The radial-velocity fitting method}

Radial velocities were determined using the {\tt iraf} package {\tt rvsao} \cite{kurtz1998} by performing cross-correlation with the task {\tt xcsao}. Barycentric corrections were calculated and applied within this same task. We initially used a spectral template for a pure-hydrogen atmosphere with $T\eff = 30\,000$~K and $\log~g = 5.5$. Each spectrum was then Doppler-corrected, and all spectra from the same arm were co-added. The co-added spectrum was used as input for a spectral fit. A template calculated using the best parameters from this spectroscopic fit was then used to re-calculate the radial velocities with {\tt xcsao}. The values obtained (given in Supplementary Tables \ref{rv_values_b} and \ref{rv_values_r}) were consistent with the initial values using the pure-hydrogen template, but uncertainties were lower by a factor of two. We performed a fit to the radial velocity curve using the Markov-chain Monte Carlo method (MCMC) implemented with the {\tt emcee} package \cite{emcee}. We assume a circular orbit, and thus fit the radial velocities using:
\begin{eqnarray}
RV (t) &=& V_0 + K\sdB \sin [2\pi (t - t_0)/P],
\end{eqnarray}
where $V_0$ is the systemic velocity, $K\sdB$ is the radial velocity semi-amplitude of the hot subdwarf, $t_0$ is the zero point of the ephemeris, and $P$ is the orbital period. The period was fixed to the photometric period, whereas $t_0$ was allowed to vary P/2. The velocities $V_0$ and $K\sdB$ were left to vary freely within physical limits, but we used values from $\chi^2$ minimisation as the initial guess. We note that, given that the spectral lines of hot subdwarfs are inherently broad, the effect of orbital smearing for our integration time of 2\% of the orbital period is of the same order of the radial velocity uncertainties, which were taken into account in our MCMC fit. We obtained $V_0 = 8.2\pm0.8$~km~s$^{-1}$ and $K\sdB = 343.1\pm1.2$~km~s$^{-1}$.

\subsection*{The Galactic orbit of HD~265435}

We calculated the Galactic orbit of HD~265435 using the package {\tt galpy} \cite{bovy2015}. The Galactic potential was modelled with three components (bulge, disk, and halo) plus a central black hole of mass $4 \times 10^6 M\sun$ \cite{bovy2013, bovy2015}. The Sun was placed at a distance of $R_0 = 8.27\pm0.29$~kpc from the Galactic centre with peculiar motion in the Local Standard of Rest of $(U\sun, V\sun, W\sun) = (11.1, 12.24, 7.25)$~km~s$^{-1}$, and the Milky way rotation speed at the Solar circle was set to $V_c = 238\pm9$~km~s$^{-1}$ \cite{schonrich2010, schonrich2012}. The system shows dynamics consistent with the thin disk of the Galaxy, as can be seen in Supplementary Fig.~\ref{orbit}. Other indicators, namely the Galactic velocity components ($U = 11\pm2$~km~s$^{-1}$, $V = 232\pm4$~km~s$^{-1}$, $W = -14\pm2$~km~s$^{-1}$), and angular momentum and eccentricity ($J_z = 2130\pm3$~kpc~km$^{-1}$~s$^{-1}$ and $e = 0.227+/-0.001$), also point to thin disk membership\cite{pauli2006}.

\subsection*{The light-curve fitting method}

We used {\sc lcurve} \cite{copperwheat2010} to carry out the light curve analysis. This code uses a grid of points to model the two stars, with shapes set by a Roche potential. The flux emitted by each grid point is calculated from a blackbody with a given estimated temperature at the bandpass wavelength, taking into account corrections for limb darkening, gravity darkening, Doppler beaming, and reflection effects.

We assume that the orbit is circular, as done for the radial velocity fit. The temperature of the hot subdwarf was fixed at the value determined from the spectroscopy. The lack of blue-excess in the SED fit implies a maximum temperature for the companion of $T\comp \approx 90\,000$~K. Assuming the minimum mass obtained from the spectroscopy and the mass-radius relationship for a carbon-oxygen white dwarf (which will give the maximum radius, and therefore maximum luminosity, for the unseen companion), we obtain that the companion contribution to the light is no more than 0.5\%. Test-runs of an MCMC fit to the light curve showed that the companion temperature and radius indeed cannot be constrained given the lack of contribution to the observed flux. Therefore we kept the companion temperature and radius fixed to arbitrary values of $T\comp = 30\,000$~K and $R\comp/a = 0.0125$, which are consistent with a white dwarf and result in a contribution of $\sim 0.15$\%.

For the hot subdwarf, limb-darkening and gravity darkening coefficients, as well as the Doppler boosting factor, were interpolated from the tables 4 and $y$ of Claret et al. 2020 \cite{claret2020}. We used the values for a pure-hydrogen composition ($\log[He/H] = -10.0$), because coefficients are unavailable at our derived values of $T\eff$ and $\log~g$ for models with intermediate helium abundance. Test-runs indicate that this choice does not affect our solution significantly, and leaving these parameters as free results in values consistent with the tabulated values that were used. These coefficients were all set to zero for the secondary, as it has no measurable contribution. This leaves as free parameters in our light curve fit the mass ratio $q$, the inclination angle $i$, the scaled radius of the primary $r\sdB = R\sdB/a$, where $a$ is the orbital distance, and the velocity scale, $V_{\textrm{scale}} = (K\sdB + K\comp)/\sin~i$.

We first computed a preliminary fit of the light curve phased to the period determined from the Lomb-Scargle periodogram using the Levenberg–Marquardt algorithm for $\chi^2$ minimisation. For this initial fit our goal was not to obtain the best physical solution, but to calculate a model describing the light curve well enough to allow a more precise determination of the ephemeris, and to subtract the binary contribution from the light curve in order to determine the contribution of pulsations. The pulsation analysis was carried out with {\sc Period04} \cite{lenz2014}. We found 33 frequencies above a detection level of five times the average amplitude of the Fourier transform. We performed a global fit in {\sc Period04} using all 33 identified frequencies, and subtracted the obtained model from the original light curve.

Next we performed a MCMC fit to the pulsations-subtracted light curve phase-folded to the determined period. The starting point were the parameters obtained from spectroscopic analyses. Initially we let all free parameters vary freely by imposing no priors on their values. This resulted on a solution with a velocity scale that was inconsistent with the observed radial velocity semi-amplitude of the hot subdwarf. We have then performed a new fit, applying a Gaussian prior to the radial velocity semi-amplitude of the hot subdwarf, to guarantee its consistency with the value estimated from spectroscopic observations. Our best-fit model is shown in the top panel of Fig.~\ref{twophase}, and the corner plot for our MCMC run is shown in Supplementary Fig.~\ref{corner}. We note that the quoted $\log~g$ for the light curve solution was computed using a flux-weighted radius, not the equatorial radius. 
\subsection*{The evolution of the system}

We conducted a numerical analysis of this system using the Modules for Experiments in Stellar Astrophysics (MESA) release 12778 \cite{MESA1,MESA2,MESA3,MESA4,MESA5}. We note that the radius of the hot subdwarf component radius is about $1.7$ times that of a pure helium hot subdwarf of the same mass \cite{heber2016}, which hints at the presence of a thin ($\leq 10^{-3}\,\text{M}_\odot$) remnant hydrogen-rich envelope, which is typical of observed hot subdwarf stars.
We found that the observational parameters best fit an evolved helium star with a remnant hydrogen envelope of $1.5 \times 10^{-4}\,\text{M}_\odot$ and solar metallicity.
This model was created by initialising a hydrogen depleted pre-MS model with solar metallicity and allowing it to contract, with nuclear burning disabled, to hydrostatic equilibrium. A hydrogen rich envelope of appropriate mass and metallicity was then accreted onto this model and the model again relaxed to hydrostatic equilibrium.
Following this, the model was evolved until it matched the observational properties of the observed hot subdwarf star. In order to preserve consistency, this was repeated for an alternative model with the inclusion of a weak wind (see below). We found that, with a wind, an initial hydrogen envelope of $3.0 \times 10^{-4}\,\text{M}_\odot$ was required. At the point where the alternative model matched the observations, about half of this envelope had been removed by winds. However, in order to evolve from its initial position in the Hertzsprung–Russell diagram, the benchmark model took $\sim 31.2\,\text{Myr}$, while the alternative model only required $\sim 26.2\,\text{Myr}$.
The initial model was then placed in a binary system with an orbital period of $P_\text{orb} = 101\,\text{min}$ and a $1.02\,\text{M}_\odot$ white dwarf, approximated as a point mass. 

In order to model the evolution of the hot subdwarf star, we use the predictive mixing scheme included in MESA with the the same parameter setting as Ostrowski et al. 2020 \cite{Ostrowski2020}, Appendix B, and a semi-convection parameter of $\alpha=0.1$. (Note: Inclusion of semi-convection tends to induce so-called "breathing pulses" in the models, which are deemed non-physical numerical artefacts. We choose the semi-convection parameter in our study such as to avoid this issue.) We note that this scheme tends to underestimate the growth of the hot subdwarf's convective core, leading to an underestimation of the hot subdwarf's remaining He-burning lifetime. This does not significantly impact the reliability of our predictions, which are in agreement with previous studies. 
As hot subdwarf winds are currently a matter of debate, we present our analysis under two different assumptions: no winds (the standard assumption) and a nominal wind following a prescription by de Jager et al. 1998 \cite{jager1988}. The latter prescription yields wind mass loss rates on the order of $10^{-11}~M\sun/\mathrm{yr}$, which agrees well with theoretical predictions \cite{Unglaub2008}, but falls short of the $10^{-9}~M\sun/\mathrm{yr}$ claimed by more recent studies \cite{Krticka2019}. 
We include the effects of rotation, assuming tidal locking and angular momentum loss through gravitational radiation, while RLOF-driven mass transfer is assumed to be conservative. 
Preceding any interaction, the system is expected to lose angular momentum through the emission of gravitational radiation, leading to RLOF after $\sim 29.6\,\text{Myr}$ (no wind). For the following $\sim 3.0\,\text{Myr}$ of this RLOF phase, the transferred material will be hydrogen enriched with mass transfer rates well below $10^{-9}\,\text{M}_\odot\,\text{yr}^{-1}$. This may be sufficient to lead to a series of classical nova outbursts by the accreting white dwarf \cite{iben1992,shara2018}. In the alternative scenario with a weak wind, RLOF is expected to occur after $\sim 37\,\text{Myr}$. In either case, the phase of He-enriched mass transfer is expected to last $\sim 1.0\,\text{Myr}$, at the end of which the envelope of the hot subdwarf will have lost sufficient amounts of He due to ongoing nuclear burning and mass transfer for continued He fusion to become unsustainable. 
At the end of He-burning in the hot subdwarf star and its subsequent contraction to a CO white dwarf, a significant amount of unburnt helium ($\sim 0.03\,\text{M}_\odot$) will remain on its surface, sufficient to classify it as a hybrid HeCO white dwarf.
We note that our simulations and predictions on the amount of remaining and transferred helium independently corroborate previous studies of systems of this type \cite{neunteufel2016,Zenati2019}.
This He-rich material will be accumulated at rates not exceeding $2.5 \times 10^{-8}~M\sun/\mathrm{yr}$, sufficiently low in order to allow for the material to be accumulated without ignition (i.e. quiescently).
Quiescent accumulation allows for building up a significant layer of unburnt helium on the white dwarf. When this pristine helium ignites explosively, ignition of the underlying carbon-oxygen core may follow, leading to a SN \cite{nomoto1980,livne1990,shen2014} (double detonation mechanism), however, the amount of material transferred is too small by $\geq 0.04~M\sun$.
With nuclear burning thus quenched, the hot subdwarf will contract to become a white dwarf. The thus formed close double degenerate binary will then merge $\sim 71.8\,\text{Myr}$ from the current epoch. Although the mass ratio remains below that previously predicted \cite{pakmor2010,pakmor2011,pakmor2012,röpke2012,sato2016} for likely progenitor systems of typical SN Ia, of $\approx1.11$, a thermonuclear supernova is the most likely outcome, given the high total mass of the system. However, the observed spectrum may not resemble a typical SN Ia, but instead a SN Iax or an otherwise subluminous or spectrally peculiar SN Ia \cite{li2003,pakmor2010,foley2013,woosley2011,wang2013}.

A number of viable channels for the production of a binaries like HD~265435 have been proposed in the past \cite{heber2016}. A hot subdwarf star of this mass is expected to form via unstable RLOF at the end of its progenitor's first giant branch (FGB), leading to  common-envelope evolution \cite{han2002,han2003}. Assuming this formation channel, using once more the MESA framework, we evolved a number of likely hot subdwarf progenitors until the end of their respective FGB. In broad agreement with Han et al. 2002, 2003 \cite{han2002,han2003} we find that the likely progenitor of the hot subdwarf is a main sequence star in in the mass range of $\sim4.3-4.4\,\text{M}_\odot$, which, with a white dwarf progenitor of $\sim 6\,\text{M}_\odot$ \cite{weidemann2000}, will determine the evolutionary timescale of the system. Note that Han et al. 2002,2003 \cite{han2002,han2003} do not provide delay times, leading to our retread of their analysis. Discrepancies are due to the utilised overshooting prescription. Under these conditions, the lifetime of the progenitor binary is $\sim 140\,\text{Myr}$. The system requires an additional $\sim 31.2\,\text{Myr}$ (benchmark) $\sim 26.2\,\text{Myr}$ (alternative) for the hot subdwarf to evolve to its currently observed physical properties. Accordingly, the timescale from zero-age main sequence to merger would be on the order of $\sim 238-243\,\text{Myr}$.

\section*{Data availability}
The TESS data used in this work are public and can be accessed via the Barbara A. Mikulski Archive for Space Telescopes (\href{https://mast.stsci.edu/}{https://mast.stsci.edu/}). Obtained follow-up spectra, evolutionary models and MESA inlists are available in Zenodo (\href{https://doi.org/10.5281/zenodo.4792303}{https://doi.org/10.5281/zenodo.4792303}).

\section*{Code availability}
The \texttt{PyRAF}-based pipeline for DBSP speectra reduction is available at \href{https://github.com/ebellm/pyraf-dbsp}{https://github.com/ebellm/pyraf-dbsp}, whereas the \texttt{MAKEE} pipeline for ESI spectra can be found at \href{http://www.astro.caltech.edu/$\sim$tb/ipac$\_$staff/tab/makee/}{http://www.astro.caltech.edu/$\sim$tb/ipac$\_$staff/tab/makee/}. 
The radial velocity determination code {\tt rvsao} is available from \href{http://tdc-www.harvard.edu/iraf/rvsao/}{http://tdc-www.harvard.edu/iraf/rvsao/}. The package {\tt galpy} can be installed following \href{https://docs.galpy.org/en/v1.6.0/}{https://docs.galpy.org/en/v1.6.0/}. The SED and spectral fitting routines are publicly documented as described above, but not publicly available. The software employed for pre-whitening the light curve, {\sc Period04}, can be obtained from \href{https://www.univie.ac.at/tops/Period04/}{https://www.univie.ac.at/tops/Period04/}. {\sc lcurve} is available at \href{https://github.com/trmrsh/cpp-lcurve}{https://github.com/trmrsh/cpp-lcurve}. The stellar evolution code {\sc mesa} can be downloaded from \href{http://mesa.sourceforge.net/}{http://mesa.sourceforge.net/}.

\bibliography{HD265435}

\begin{thebibliography}{100}
\urlstyle{rm}
\expandafter\ifx\csname url\endcsname\relax
  \def\url#1{\texttt{#1}}\fi
\expandafter\ifx\csname urlprefix\endcsname\relax\def\urlprefix{URL }\fi
\expandafter\ifx\csname doiprefix\endcsname\relax\def\doiprefix{DOI: }\fi
\providecommand{\bibinfo}[2]{#2}
\providecommand{\eprint}[2][]{\url{#2}}

\bibitem{schmidt1998}
\bibinfo{author}{{Schmidt}, B.~P.} et~al.
\newblock \bibinfo{journal}{\bibinfo{title}{{The High-Z Supernova Search:
  Measuring Cosmic Deceleration and Global Curvature of the Universe Using Type
  IA Supernovae}}}.
\newblock {\it {\JournalTitle{\apj}}} \textbf{\bibinfo{volume}{507}},
  \bibinfo{pages}{46--63} (\bibinfo{year}{1998}).

\bibitem{riess1998}
\bibinfo{author}{{Riess}, A.~G.} et~al.
\newblock \bibinfo{journal}{\bibinfo{title}{{Observational Evidence from
  Supernovae for an Accelerating Universe and a Cosmological Constant}}}.
\newblock {\it {\JournalTitle{\aj}}} \textbf{\bibinfo{volume}{116}},
  \bibinfo{pages}{1009--1038} (\bibinfo{year}{1998}).

\bibitem{perlmutter1999}
\bibinfo{author}{{Perlmutter}, S.} et~al.
\newblock \bibinfo{journal}{\bibinfo{title}{{Measurements of
  {\ensuremath{\Omega}} and {\ensuremath{\Lambda}} from 42 High-Redshift
  Supernovae}}}.
\newblock {\it {\JournalTitle{\apj}}} \textbf{\bibinfo{volume}{517}},
  \bibinfo{pages}{565--586} (\bibinfo{year}{1999}).

\bibitem{riess2019}
\bibinfo{author}{{Riess}, A.~G.}, \bibinfo{author}{{Casertano}, S.},
  \bibinfo{author}{{Yuan}, W.}, \bibinfo{author}{{Macri}, L.~M.} \&
  \bibinfo{author}{{Scolnic}, D.}
\newblock \bibinfo{journal}{\bibinfo{title}{{Large Magellanic Cloud Cepheid
  Standards Provide a 1\% Foundation for the Determination of the Hubble
  Constant and Stronger Evidence for Physics beyond
  {\ensuremath{\Lambda}}CDM}}}.
\newblock {\it {\JournalTitle{\apj}}} \textbf{\bibinfo{volume}{876}},
  \bibinfo{pages}{85} (\bibinfo{year}{2019}).

\bibitem{planck2018}
\bibinfo{author}{{Planck Collaboration}} et~al.
\newblock \bibinfo{journal}{\bibinfo{title}{{Planck 2018 results. VI.
  Cosmological parameters}}}.
\newblock {\it {\JournalTitle{\aap}}} \textbf{\bibinfo{volume}{641}},
  \bibinfo{pages}{A6} (\bibinfo{year}{2020}).

\bibitem{bernal2016}
\bibinfo{author}{{Bernal}, J.~L.}, \bibinfo{author}{{Verde}, L.} \&
  \bibinfo{author}{{Riess}, A.~G.}
\newblock \bibinfo{journal}{\bibinfo{title}{{The trouble with H$_{0}$}}}.
\newblock {\it {\JournalTitle{\jcap}}} \textbf{\bibinfo{volume}{2016}},
  \bibinfo{pages}{019} (\bibinfo{year}{2016}).

\bibitem{hoyle1960}
\bibinfo{author}{{Hoyle}, F.} \& \bibinfo{author}{{Fowler}, W.~A.}
\newblock \bibinfo{journal}{\bibinfo{title}{{Nucleosynthesis in Supernovae.}}}
\newblock {\it {\JournalTitle{\apj}}} \textbf{\bibinfo{volume}{132}},
  \bibinfo{pages}{565} (\bibinfo{year}{1960}).

\bibitem{hillebrandt2013}
\bibinfo{author}{{Hillebrandt}, W.}, \bibinfo{author}{{Kromer}, M.},
  \bibinfo{author}{{R{\"o}pke}, F.~K.} \& \bibinfo{author}{{Ruiter}, A.~J.}
\newblock \bibinfo{journal}{\bibinfo{title}{{Towards an understanding of Type
  Ia supernovae from a synthesis of theory and observations}}}.
\newblock {\it {\JournalTitle{Frontiers of Physics}}}
  \textbf{\bibinfo{volume}{8}}, \bibinfo{pages}{116--143}
  (\bibinfo{year}{2013}).

\bibitem{whelan1973}
\bibinfo{author}{{Whelan}, J.} \& \bibinfo{author}{{Iben}, J., Icko}.
\newblock \bibinfo{journal}{\bibinfo{title}{{Binaries and Supernovae of Type
  I}}}.
\newblock {\it {\JournalTitle{\apj}}} \textbf{\bibinfo{volume}{186}},
  \bibinfo{pages}{1007--1014} (\bibinfo{year}{1973}).

\bibitem{iben1984}
\bibinfo{author}{{Iben}, J., I.} \& \bibinfo{author}{{Tutukov}, A.~V.}
\newblock \bibinfo{journal}{\bibinfo{title}{{Supernovae of type I as end
  products of the evolution of binaries with components of moderate initial
  mass.}}}
\newblock {\it {\JournalTitle{\apjs}}} \textbf{\bibinfo{volume}{54}},
  \bibinfo{pages}{335--372} (\bibinfo{year}{1984}).

\bibitem{liu2018}
\bibinfo{author}{{Liu}, D.}, \bibinfo{author}{{Wang}, B.} \&
  \bibinfo{author}{{Han}, Z.}
\newblock \bibinfo{journal}{\bibinfo{title}{{The double-degenerate model for
  the progenitors of Type Ia supernovae}}}.
\newblock {\it {\JournalTitle{\mnras}}} \textbf{\bibinfo{volume}{473}},
  \bibinfo{pages}{5352--5361} (\bibinfo{year}{2018}).

\bibitem{han2004}
\bibinfo{author}{{Han}, Z.} \& \bibinfo{author}{{Podsiadlowski}, P.}
\newblock \bibinfo{journal}{\bibinfo{title}{{The single-degenerate channel for
  the progenitors of Type Ia supernovae}}}.
\newblock {\it {\JournalTitle{\mnras}}} \textbf{\bibinfo{volume}{350}},
  \bibinfo{pages}{1301--1309} (\bibinfo{year}{2004}).

\bibitem{rebassa2019}
\bibinfo{author}{{Rebassa-Mansergas}, A.}, \bibinfo{author}{{Toonen}, S.},
  \bibinfo{author}{{Korol}, V.} \& \bibinfo{author}{{Torres}, S.}
\newblock \bibinfo{journal}{\bibinfo{title}{{Where are the double-degenerate
  progenitors of Type Ia supernovae?}}}
\newblock {\it {\JournalTitle{\mnras}}} \textbf{\bibinfo{volume}{482}},
  \bibinfo{pages}{3656--3668} (\bibinfo{year}{2019}).

\bibitem{maoz2012}
\bibinfo{author}{{Maoz}, D.} \& \bibinfo{author}{{Mannucci}, F.}
\newblock \bibinfo{journal}{\bibinfo{title}{{Type-Ia Supernova Rates and the
  Progenitor Problem: A Review}}}.
\newblock {\it {\JournalTitle{\pasa}}} \textbf{\bibinfo{volume}{29}},
  \bibinfo{pages}{447--465} (\bibinfo{year}{2012}).

\bibitem{santander2015}
\bibinfo{author}{{Santander-Garc{\'\i}a}, M.} et~al.
\newblock \bibinfo{journal}{\bibinfo{title}{{The double-degenerate,
  super-Chandrasekhar nucleus of the planetary nebula Henize 2-428}}}.
\newblock {\it {\JournalTitle{\nat}}} \textbf{\bibinfo{volume}{519}},
  \bibinfo{pages}{63--65} (\bibinfo{year}{2015}).

\bibitem{reindl2020}
\bibinfo{author}{{Reindl}, N.} et~al.
\newblock \bibinfo{journal}{\bibinfo{title}{{An in-depth reanalysis of the
  alleged type Ia supernova progenitor Henize 2-428}}}.
\newblock {\it {\JournalTitle{\aap}}} \textbf{\bibinfo{volume}{638}},
  \bibinfo{pages}{A93} (\bibinfo{year}{2020}).

\bibitem{napiwotzki2020}
\bibinfo{author}{{Napiwotzki}, R.} et~al.
\newblock \bibinfo{journal}{\bibinfo{title}{{The ESO supernovae type Ia
  progenitor survey (SPY). The radial velocities of 643 DA white dwarfs}}}.
\newblock {\it {\JournalTitle{\aap}}} \textbf{\bibinfo{volume}{638}},
  \bibinfo{pages}{A131} (\bibinfo{year}{2020}).

\bibitem{maxted2000}
\bibinfo{author}{{Maxted}, P.~F.~L.}, \bibinfo{author}{{Marsh}, T.~R.} \&
  \bibinfo{author}{{North}, R.~C.}
\newblock \bibinfo{journal}{\bibinfo{title}{{KPD 1930+2752: a candidate Type Ia
  supernova progenitor}}}.
\newblock {\it {\JournalTitle{\mnras}}} \textbf{\bibinfo{volume}{317}},
  \bibinfo{pages}{L41--L44} (\bibinfo{year}{2000}).

\bibitem{vennes2012}
\bibinfo{author}{{Vennes}, S.}, \bibinfo{author}{{Kawka}, A.},
  \bibinfo{author}{{O'Toole}, S.~J.}, \bibinfo{author}{{N{\'e}meth}, P.} \&
  \bibinfo{author}{{Burton}, D.}
\newblock \bibinfo{journal}{\bibinfo{title}{{The Shortest Period sdB Plus White
  Dwarf Binary CD-30 11223 (GALEX J1411-3053)}}}.
\newblock {\it {\JournalTitle{\apjl}}} \textbf{\bibinfo{volume}{759}},
  \bibinfo{pages}{L25} (\bibinfo{year}{2012}).

\bibitem{geier2013}
\bibinfo{author}{{Geier}, S.} et~al.
\newblock \bibinfo{journal}{\bibinfo{title}{{A progenitor binary and an ejected
  mass donor remnant of faint type Ia supernovae}}}.
\newblock {\it {\JournalTitle{\aap}}} \textbf{\bibinfo{volume}{554}},
  \bibinfo{pages}{A54} (\bibinfo{year}{2013}).

\bibitem{ricker2015}
\bibinfo{author}{{Ricker}, G.~R.} et~al.
\newblock \bibinfo{journal}{\bibinfo{title}{{Transiting Exoplanet Survey
  Satellite (TESS)}}}.
\newblock {\it {\JournalTitle{Journal of Astronomical Telescopes, Instruments,
  and Systems}}} \textbf{\bibinfo{volume}{1}}, \bibinfo{pages}{014003}
  (\bibinfo{year}{2015}).

\bibitem{oke1982}
\bibinfo{author}{{Oke}, J.~B.} \& \bibinfo{author}{{Gunn}, J.~E.}
\newblock \bibinfo{journal}{\bibinfo{title}{{An Efficient Low Resolution and
  Moderate Resolution Spectrograph for the Hale Telescope}}}.
\newblock {\it {\JournalTitle{\pasp}}} \textbf{\bibinfo{volume}{94}},
  \bibinfo{pages}{586} (\bibinfo{year}{1982}).

\bibitem{gaia_edr3}
\bibinfo{author}{{Gaia Collaboration}} et~al.
\newblock \bibinfo{journal}{\bibinfo{title}{{Gaia Early Data Release 3. Summary
  of the contents and survey properties}}}.
\newblock {\it {\JournalTitle{\aap}}} \textbf{\bibinfo{volume}{649}},
  \bibinfo{pages}{A1} (\bibinfo{year}{2021}).

\bibitem{edr3_zeropoint}
\bibinfo{author}{{Lindegren}, L.} et~al.
\newblock \bibinfo{journal}{\bibinfo{title}{{Gaia Early Data Release 3.
  Parallax bias versus magnitude, colour, and position}}}.
\newblock {\it {\JournalTitle{\aap}}} \textbf{\bibinfo{volume}{649}},
  \bibinfo{pages}{A4} (\bibinfo{year}{2021}).

\bibitem{shakura1987}
\bibinfo{author}{{Shakura}, N.~I.} \& \bibinfo{author}{{Postnov}, K.~A.}
\newblock \bibinfo{journal}{\bibinfo{title}{{Doppler-effect modulation of the
  observed radiation flux from ultracompact binary stars.}}}
\newblock {\it {\JournalTitle{\aap}}} \textbf{\bibinfo{volume}{183}},
  \bibinfo{pages}{L21--L22} (\bibinfo{year}{1987}).

\bibitem{charpinet1996}
\bibinfo{author}{{Charpinet}, S.}, \bibinfo{author}{{Fontaine}, G.},
  \bibinfo{author}{{Brassard}, P.} \& \bibinfo{author}{{Dorman}, B.}
\newblock \bibinfo{journal}{\bibinfo{title}{{The Potential of Asteroseismology
  for Hot, Subdwarf B Stars: A New Class of Pulsating Stars?}}}
\newblock {\it {\JournalTitle{\apjl}}} \textbf{\bibinfo{volume}{471}},
  \bibinfo{pages}{L103} (\bibinfo{year}{1996}).

\bibitem{kilkenny1997}
\bibinfo{author}{{Kilkenny}, D.}, \bibinfo{author}{{Koen}, C.},
  \bibinfo{author}{{O'Donoghue}, D.} \& \bibinfo{author}{{Stobie}, R.~S.}
\newblock \bibinfo{journal}{\bibinfo{title}{{A new class of rapidly pulsating
  star - I. EC 14026-2647, the class prototype}}}.
\newblock {\it {\JournalTitle{\mnras}}} \textbf{\bibinfo{volume}{285}},
  \bibinfo{pages}{640--644} (\bibinfo{year}{1997}).

\bibitem{kawaler2005}
\bibinfo{author}{{Kawaler}, S.~D.} \& \bibinfo{author}{{Hostler}, S.~R.}
\newblock \bibinfo{journal}{\bibinfo{title}{{Internal Rotation of Subdwarf B
  Stars: Limiting Cases and Asteroseismological Consequences}}}.
\newblock {\it {\JournalTitle{\apj}}} \textbf{\bibinfo{volume}{621}},
  \bibinfo{pages}{432--444} (\bibinfo{year}{2005}).

\bibitem{reed2014}
\bibinfo{author}{{Reed}, M.~D.} et~al.
\newblock \bibinfo{journal}{\bibinfo{title}{{Analysis of the rich frequency
  spectrum of KIC 10670103 revealing the most slowly rotating subdwarf B star
  in the Kepler field}}}.
\newblock {\it {\JournalTitle{\mnras}}} \textbf{\bibinfo{volume}{440}},
  \bibinfo{pages}{3809--3824} (\bibinfo{year}{2014}).

\bibitem{geier2008}
\bibinfo{author}{{Geier}, S.}, \bibinfo{author}{{Karl}, C.},
  \bibinfo{author}{{Edelmann}, H.}, \bibinfo{author}{{Heber}, U.} \&
  \bibinfo{author}{{Napiwotzki}, R.}
\newblock \bibinfo{title}{{Binary sdB Stars with Massive Compact Companions}}.
\newblock In \bibinfo{editor}{{Heber}, U.}, \bibinfo{editor}{{Jeffery}, C.~S.}
  \& \bibinfo{editor}{{Napiwotzki}, R.} (eds.) {\bibinfo{booktitle}{Hot
  Subdwarf Stars and Related Objects}}, vol. \bibinfo{volume}{392} of
  {\bibinfo{series}{Astronomical Society of the Pacific Conference Series}},
  \bibinfo{pages}{207} (\bibinfo{year}{2008}).

\bibitem{geier2019}
\bibinfo{author}{{Geier}, S.}, \bibinfo{author}{{Raddi}, R.},
  \bibinfo{author}{{Gentile Fusillo}, N.~P.} \& \bibinfo{author}{{Marsh},
  T.~R.}
\newblock \bibinfo{journal}{\bibinfo{title}{{The population of hot subdwarf
  stars studied with Gaia. II. The Gaia DR2 catalogue of hot subluminous
  stars}}}.
\newblock {\it {\JournalTitle{\aap}}} \textbf{\bibinfo{volume}{621}},
  \bibinfo{pages}{A38} (\bibinfo{year}{2019}).

\bibitem{copperwheat2010}
\bibinfo{author}{{Copperwheat}, C.~M.} et~al.
\newblock \bibinfo{journal}{\bibinfo{title}{{Physical properties of IP Pegasi:
  an eclipsing dwarf nova with an unusually cool white dwarf}}}.
\newblock {\it {\JournalTitle{\mnras}}} \textbf{\bibinfo{volume}{402}},
  \bibinfo{pages}{1824--1840} (\bibinfo{year}{2010}).

\bibitem{lauffer2018}
\bibinfo{author}{{Lauffer}, G.~R.}, \bibinfo{author}{{Romero}, A.~D.} \&
  \bibinfo{author}{{Kepler}, S.~O.}
\newblock \bibinfo{journal}{\bibinfo{title}{{New full evolutionary sequences of
  H- and He-atmosphere massive white dwarf stars using MESA}}}.
\newblock {\it {\JournalTitle{\mnras}}} \textbf{\bibinfo{volume}{480}},
  \bibinfo{pages}{1547--1562} (\bibinfo{year}{2018}).

\bibitem{heber2016}
\bibinfo{author}{{Heber}, U.}
\newblock \bibinfo{journal}{\bibinfo{title}{{Hot Subluminous Stars}}}.
\newblock {\it {\JournalTitle{\pasp}}} \textbf{\bibinfo{volume}{128}},
  \bibinfo{pages}{082001} (\bibinfo{year}{2016}).

\bibitem{Unglaub2008}
\bibinfo{author}{{Unglaub}, K.}
\newblock \bibinfo{journal}{\bibinfo{title}{{Mass-loss and diffusion in
  subdwarf B stars and hot white dwarfs: do weak winds exist?}}}
\newblock {\it {\JournalTitle{\aap}}} \textbf{\bibinfo{volume}{486}},
  \bibinfo{pages}{923--940} (\bibinfo{year}{2008}).

\bibitem{iben1992}
\bibinfo{author}{{Iben}, J., Icko}, \bibinfo{author}{{Fujimoto}, M.~Y.} \&
  \bibinfo{author}{{MacDonald}, J.}
\newblock \bibinfo{journal}{\bibinfo{title}{{On Mass-Transfer Rates in
  Classical Nova Precursors}}}.
\newblock {\it {\JournalTitle{\apj}}} \textbf{\bibinfo{volume}{384}},
  \bibinfo{pages}{580} (\bibinfo{year}{1992}).

\bibitem{shara2018}
\bibinfo{author}{{Shara}, M.~M.}, \bibinfo{author}{{Prialnik}, D.},
  \bibinfo{author}{{Hillman}, Y.} \& \bibinfo{author}{{Kovetz}, A.}
\newblock \bibinfo{journal}{\bibinfo{title}{{The Masses and Accretion Rates of
  White Dwarfs in Classical and Recurrent Novae}}}.
\newblock {\it {\JournalTitle{\apj}}} \textbf{\bibinfo{volume}{860}},
  \bibinfo{pages}{110} (\bibinfo{year}{2018}).

\bibitem{woosley2011}
\bibinfo{author}{{Woosley}, S.~E.} \& \bibinfo{author}{{Kasen}, D.}
\newblock \bibinfo{journal}{\bibinfo{title}{{Sub-Chandrasekhar Mass Models for
  Supernovae}}}.
\newblock {\it {\JournalTitle{\apj}}} \textbf{\bibinfo{volume}{734}},
  \bibinfo{pages}{38} (\bibinfo{year}{2011}).

\bibitem{neunteufel2016}
\bibinfo{author}{{Neunteufel}, P.}, \bibinfo{author}{{Yoon}, S.~C.} \&
  \bibinfo{author}{{Langer}, N.}
\newblock \bibinfo{journal}{\bibinfo{title}{{Models for the evolution of close
  binaries with He-star and white dwarf components towards Type Ia supernova
  explosions}}}.
\newblock {\it {\JournalTitle{\aap}}} \textbf{\bibinfo{volume}{589}},
  \bibinfo{pages}{A43} (\bibinfo{year}{2016}).

\bibitem{Tutukov1979}
\bibinfo{author}{{Tutukov}, A.~V.} \& \bibinfo{author}{{Yungelson}, L.~R.}
\newblock \bibinfo{journal}{\bibinfo{title}{{On the influence of emission of
  gravitational waves on the evolution of low-mass close binary stars.}}}
\newblock {\it {\JournalTitle{\actaa}}} \textbf{\bibinfo{volume}{29}},
  \bibinfo{pages}{665--680} (\bibinfo{year}{1979}).

\bibitem{neunteufel2020}
\bibinfo{author}{{Neunteufel}, P.}
\newblock \bibinfo{journal}{\bibinfo{title}{{Exploring velocity limits in the
  thermonuclear supernova ejection scenario for hypervelocity stars and the
  origin of US 708}}}.
\newblock {\it {\JournalTitle{\aap}}} \textbf{\bibinfo{volume}{641}},
  \bibinfo{pages}{A52} (\bibinfo{year}{2020}).

\bibitem{Kromer2010}
\bibinfo{author}{{Kromer}, M.} et~al.
\newblock \bibinfo{journal}{\bibinfo{title}{{Double-detonation
  Sub-Chandrasekhar Supernovae: Synthetic Observables for Minimum Helium Shell
  Mass Models}}}.
\newblock {\it {\JournalTitle{\apj}}} \textbf{\bibinfo{volume}{719}},
  \bibinfo{pages}{1067--1082} (\bibinfo{year}{2010}).

\bibitem{pakmor2010}
\bibinfo{author}{{Pakmor}, R.} et~al.
\newblock \bibinfo{journal}{\bibinfo{title}{{Sub-luminous type Ia supernovae
  from the mergers of equal-mass white dwarfs with mass
  \raisebox{-0.5ex}\textasciitilde0.9M$_{solar}$}}}.
\newblock {\it {\JournalTitle{\nat}}} \textbf{\bibinfo{volume}{463}},
  \bibinfo{pages}{61--64} (\bibinfo{year}{2010}).

\bibitem{pakmor2021}
\bibinfo{author}{{Pakmor}, R.}, \bibinfo{author}{{Zenati}, Y.},
  \bibinfo{author}{{Perets}, H.~B.} \& \bibinfo{author}{{Toonen}, S.}
\newblock \bibinfo{journal}{\bibinfo{title}{{Thermonuclear explosion of a
  massive hybrid HeCO white dwarf triggered by a He detonation on a
  companion}}}.
\newblock {\it {\JournalTitle{\mnras}}} \textbf{\bibinfo{volume}{503}},
  \bibinfo{pages}{4734--4747} (\bibinfo{year}{2021}).

\bibitem{kraft1962}
\bibinfo{author}{{Kraft}, R.~P.}, \bibinfo{author}{{Mathews}, J.} \&
  \bibinfo{author}{{Greenstein}, J.~L.}
\newblock \bibinfo{journal}{\bibinfo{title}{{Binary Stars among Cataclysmic
  Variables. II. Nova WZ Sagittae: a Possible Radiator of Gravitational
  Waves.}}}
\newblock {\it {\JournalTitle{\apj}}} \textbf{\bibinfo{volume}{136}},
  \bibinfo{pages}{312--315} (\bibinfo{year}{1962}).

\bibitem{shah2012}
\bibinfo{author}{{Shah}, S.}, \bibinfo{author}{{van der Sluys}, M.} \&
  \bibinfo{author}{{Nelemans}, G.}
\newblock \bibinfo{journal}{\bibinfo{title}{{Using electromagnetic observations
  to aid gravitational-wave parameter estimation of compact binaries observed
  with LISA}}}.
\newblock {\it {\JournalTitle{\aap}}} \textbf{\bibinfo{volume}{544}},
  \bibinfo{pages}{A153} (\bibinfo{year}{2012}).

\bibitem{moore2015}
\bibinfo{author}{{Moore}, C.~J.}, \bibinfo{author}{{Cole}, R.~H.} \&
  \bibinfo{author}{{Berry}, C.~P.~L.}
\newblock \bibinfo{journal}{\bibinfo{title}{{Gravitational-wave sensitivity
  curves}}}.
\newblock {\it {\JournalTitle{Classical and Quantum Gravity}}}
  \textbf{\bibinfo{volume}{32}}, \bibinfo{pages}{015014}
  (\bibinfo{year}{2015}).

\bibitem{gronow2020}
\bibinfo{author}{{Gronow}, S.} et~al.
\newblock \bibinfo{journal}{\bibinfo{title}{{SNe Ia from double detonations:
  Impact of core-shell mixing on the carbon ignition mechanism}}}.
\newblock {\it {\JournalTitle{\aap}}} \textbf{\bibinfo{volume}{635}},
  \bibinfo{pages}{A169} (\bibinfo{year}{2020}).

\bibitem{robson2019}
\bibinfo{author}{{Robson}, T.}, \bibinfo{author}{{Cornish}, N.~J.} \&
  \bibinfo{author}{{Liu}, C.}
\newblock \bibinfo{journal}{\bibinfo{title}{{The construction and use of LISA
  sensitivity curves}}}.
\newblock {\it {\JournalTitle{Classical and Quantum Gravity}}}
  \textbf{\bibinfo{volume}{36}}, \bibinfo{pages}{105011}
  (\bibinfo{year}{2019}).

\bibitem{kupfer2018}
\bibinfo{author}{{Kupfer}, T.} et~al.
\newblock \bibinfo{journal}{\bibinfo{title}{{LISA verification binaries with
  updated distances from Gaia Data Release 2}}}.
\newblock {\it {\JournalTitle{\mnras}}} \textbf{\bibinfo{volume}{480}},
  \bibinfo{pages}{302--309} (\bibinfo{year}{2018}).

\bibitem{kupfer2020b}
\bibinfo{author}{{Kupfer}, T.} et~al.
\newblock \bibinfo{journal}{\bibinfo{title}{{A New Class of Roche Lobe-filling
  Hot Subdwarf Binaries}}}.
\newblock {\it {\JournalTitle{\apjl}}} \textbf{\bibinfo{volume}{898}},
  \bibinfo{pages}{L25} (\bibinfo{year}{2020}).

\bibitem{juric2008}
\bibinfo{author}{{Juri{\'c}}, M.} et~al.
\newblock \bibinfo{journal}{\bibinfo{title}{{The Milky Way Tomography with
  SDSS. I. Stellar Number Density Distribution}}}.
\newblock {\it {\JournalTitle{\apj}}} \textbf{\bibinfo{volume}{673}},
  \bibinfo{pages}{864--914} (\bibinfo{year}{2008}).

\bibitem{han2003}
\bibinfo{author}{{Han}, Z.}, \bibinfo{author}{{Podsiadlowski}, P.},
  \bibinfo{author}{{Maxted}, P.~F.~L.} \& \bibinfo{author}{{Marsh}, T.~R.}
\newblock \bibinfo{journal}{\bibinfo{title}{{The origin of subdwarf B stars -
  II}}}.
\newblock {\it {\JournalTitle{\mnras}}} \textbf{\bibinfo{volume}{341}},
  \bibinfo{pages}{669--691} (\bibinfo{year}{2003}).

\bibitem{wang2010}
\bibinfo{author}{{Wang}, B.} et~al.
\newblock \bibinfo{journal}{\bibinfo{title}{{Birthrates and delay times of Type
  Ia supernovae}}}.
\newblock {\it {\JournalTitle{Science China Physics, Mechanics, and
  Astronomy}}} \textbf{\bibinfo{volume}{53}}, \bibinfo{pages}{586--590}
  (\bibinfo{year}{2010}).

\bibitem{li2011}
\bibinfo{author}{{Li}, W.} et~al.
\newblock \bibinfo{journal}{\bibinfo{title}{{Nearby supernova rates from the
  Lick Observatory Supernova Search - III. The rate-size relation, and the
  rates as a function of galaxy Hubble type and colour}}}.
\newblock {\it {\JournalTitle{\mnras}}} \textbf{\bibinfo{volume}{412}},
  \bibinfo{pages}{1473--1507} (\bibinfo{year}{2011}).

\bibitem{maoz2014}
\bibinfo{author}{{Maoz}, D.}, \bibinfo{author}{{Mannucci}, F.} \&
  \bibinfo{author}{{Nelemans}, G.}
\newblock \bibinfo{journal}{\bibinfo{title}{{Observational Clues to the
  Progenitors of Type Ia Supernovae}}}.
\newblock {\it {\JournalTitle{\araa}}} \textbf{\bibinfo{volume}{52}},
  \bibinfo{pages}{107--170} (\bibinfo{year}{2014}).

\bibitem{andrae2018}
\bibinfo{author}{{Andrae}, R.} et~al.
\newblock \bibinfo{journal}{\bibinfo{title}{{Gaia Data Release 2. First stellar
  parameters from Apsis}}}.
\newblock {\it {\JournalTitle{\aap}}} \textbf{\bibinfo{volume}{616}},
  \bibinfo{pages}{A8} (\bibinfo{year}{2018}).

\bibitem{bellm2016}
\bibinfo{author}{{Bellm}, E.~C.} \& \bibinfo{author}{{Sesar}, B.}
\newblock \bibinfo{title}{{pyraf-dbsp: Reduction pipeline for the Palomar
  Double Beam Spectrograph}} (\bibinfo{year}{2016}).

\bibitem{2014A&A...565A..63I}
\bibinfo{author}{{Irrgang}, A.} et~al.
\newblock \bibinfo{journal}{\bibinfo{title}{{A new method for an objective,
  {\ensuremath{\chi}}$^{2}$-based spectroscopic analysis of early-type stars.
  First results from its application to single and binary B- and late O-type
  stars}}}.
\newblock {\it {\JournalTitle{\aap}}} \textbf{\bibinfo{volume}{565}},
  \bibinfo{pages}{A63} (\bibinfo{year}{2014}).

\bibitem{1996ASPC..108..160K}
\bibinfo{author}{{Kurucz}, R.~L.}
\newblock \bibinfo{title}{{Status of the ATLAS 12 Opacity Sampling Program and
  of New Programs for Rosseland and for Distribution Function Opacity}}.
\newblock In \bibinfo{editor}{{Adelman}, S.~J.}, \bibinfo{editor}{{Kupka}, F.}
  \& \bibinfo{editor}{{Weiss}, W.~W.} (eds.) {\bibinfo{booktitle}{M.A.S.S.,
  Model Atmospheres and Spectrum Synthesis}}, vol. \bibinfo{volume}{108} of
  {\bibinfo{series}{Astronomical Society of the Pacific Conference Series}},
  \bibinfo{pages}{160} (\bibinfo{year}{1996}).

\bibitem{1981PhDT.......113G}
\bibinfo{author}{{Giddings}, J.~R.}
\newblock Ph.D. thesis, \bibinfo{school}{Univ. London} (\bibinfo{year}{1981}).

\bibitem{detailsurface2}
\bibinfo{journal}{\bibinfo{author}{{Butler}, K.} \&
  \bibinfo{author}{{Giddings}, J.~R.}}
\newblock {\it {\JournalTitle{{Newsletter of Analysis of Astronomical
  Spectra}}}} \textbf{\bibinfo{volume}{9}} (\bibinfo{year}{1985}).

\bibitem{2018A&A...615L...5I}
\bibinfo{author}{{Irrgang}, A.}, \bibinfo{author}{{Kreuzer}, S.},
  \bibinfo{author}{{Heber}, U.} \& \bibinfo{author}{{Brown}, W.}
\newblock \bibinfo{journal}{\bibinfo{title}{{A quantitative spectral analysis
  of 14 hypervelocity stars from the MMT survey}}}.
\newblock {\it {\JournalTitle{\aap}}} \textbf{\bibinfo{volume}{615}},
  \bibinfo{pages}{L5} (\bibinfo{year}{2018}).

\bibitem{przybilla2011}
\bibinfo{author}{{Przybilla}, N.}, \bibinfo{author}{{Nieva}, M.-F.} \&
  \bibinfo{author}{{Butler}, K.}
\newblock \bibinfo{title}{{Testing common classical LTE and NLTE model
  atmosphere and line-formation codes for quantitative spectroscopy of
  early-type stars}}.
\newblock In {\bibinfo{booktitle}{Journal of Physics Conference Series}}, vol.
  \bibinfo{volume}{328} of {\bibinfo{series}{Journal of Physics Conference
  Series}}, \bibinfo{pages}{012015} (\bibinfo{year}{2011}).

\bibitem{silvotti2019}
\bibinfo{author}{{Silvotti}, R.} et~al.
\newblock \bibinfo{journal}{\bibinfo{title}{{High-degree gravity modes in the
  single sdB star HD 4539}}}.
\newblock {\it {\JournalTitle{\mnras}}} \textbf{\bibinfo{volume}{489}},
  \bibinfo{pages}{4791--4801} (\bibinfo{year}{2019}).

\bibitem{sahoo2020}
\bibinfo{author}{{Sahoo}, S.~K.} et~al.
\newblock \bibinfo{journal}{\bibinfo{title}{{Mode identification in three
  pulsating hot subdwarfs observed with TESS satellite}}}.
\newblock {\it {\JournalTitle{\mnras}}} \textbf{\bibinfo{volume}{495}},
  \bibinfo{pages}{2844--2857} (\bibinfo{year}{2020}).

\bibitem{silvotti2021}
\bibinfo{author}{{Silvotti}, R.} et~al.
\newblock \bibinfo{journal}{\bibinfo{title}{{EPIC 216747137: a new HW Vir
  eclipsing binary with a massive sdOB primary and a low-mass M-dwarf
  companion}}}.
\newblock {\it {\JournalTitle{\mnras}}} \textbf{\bibinfo{volume}{500}},
  \bibinfo{pages}{2461--2474} (\bibinfo{year}{2021}).

\bibitem{1994A&A...282..151H}
\bibinfo{author}{{Hubeny}, I.}, \bibinfo{author}{{Hummer}, D.~G.} \&
  \bibinfo{author}{{Lanz}, T.}
\newblock \bibinfo{journal}{\bibinfo{title}{{NLTE model stellar atmospheres
  with line blanketing near the series limits.}}}
\newblock {\it {\JournalTitle{\aap}}} \textbf{\bibinfo{volume}{282}},
  \bibinfo{pages}{151--167} (\bibinfo{year}{1994}).

\bibitem{2009ApJ...696.1755T}
\bibinfo{author}{{Tremblay}, P.~E.} \& \bibinfo{author}{{Bergeron}, P.}
\newblock \bibinfo{journal}{\bibinfo{title}{{Spectroscopic Analysis of DA White
  Dwarfs: Stark Broadening of Hydrogen Lines Including Nonideal Effects}}}.
\newblock {\it {\JournalTitle{\apj}}} \textbf{\bibinfo{volume}{696}},
  \bibinfo{pages}{1755--1770} (\bibinfo{year}{2009}).

\bibitem{1997ApJS..108..559B}
\bibinfo{author}{{Beauchamp}, A.}, \bibinfo{author}{{Wesemael}, F.} \&
  \bibinfo{author}{{Bergeron}, P.}
\newblock \bibinfo{journal}{\bibinfo{title}{{Spectroscopic Studies of DB White
  Dwarfs: Improved Stark Profiles for Optical Transitions of Neutral Helium}}}.
\newblock {\it {\JournalTitle{\apjs}}} \textbf{\bibinfo{volume}{108}},
  \bibinfo{pages}{559--573} (\bibinfo{year}{1997}).

\bibitem{lei2018}
\bibinfo{author}{{Lei}, Z.}, \bibinfo{author}{{Zhao}, J.},
  \bibinfo{author}{{N{\'e}meth}, P.} \& \bibinfo{author}{{Zhao}, G.}
\newblock \bibinfo{journal}{\bibinfo{title}{{New Hot Subdwarf Stars Identified
  in Gaia DR2 with LAMOST DR5 Spectra}}}.
\newblock {\it {\JournalTitle{\apj}}} \textbf{\bibinfo{volume}{868}},
  \bibinfo{pages}{70} (\bibinfo{year}{2018}).

\bibitem{2018OAst...27...35H}
\bibinfo{author}{{Heber}, U.}, \bibinfo{author}{{Irrgang}, A.} \&
  \bibinfo{author}{{Schaffenroth}, J.}
\newblock \bibinfo{journal}{\bibinfo{title}{{Spectral energy distributions and
  colours of hot subluminous stars}}}.
\newblock {\it {\JournalTitle{Open Astronomy}}} \textbf{\bibinfo{volume}{27}},
  \bibinfo{pages}{35--43} (\bibinfo{year}{2018}).

\bibitem{RUWE}
\bibinfo{author}{Lindegren, L.}
\newblock \bibinfo{title}{{R}e-normalising the astrometric chi-square in {G}aia
  {D}{R}2} (\bibinfo{year}{2018}).
\newblock \bibinfo{note}{GAIA-C3-TN-LU-LL-124}.

\bibitem{2019ApJ...886..108F}
\bibinfo{author}{{Fitzpatrick}, E.~L.}, \bibinfo{author}{{Massa}, D.},
  \bibinfo{author}{{Gordon}, K.~D.}, \bibinfo{author}{{Bohlin}, R.} \&
  \bibinfo{author}{{Clayton}, G.~C.}
\newblock \bibinfo{journal}{\bibinfo{title}{{An Analysis of the Shapes of
  Interstellar Extinction Curves. VII. Milky Way Spectrophotometric
  Optical-through-ultraviolet Extinction and Its R-dependence}}}.
\newblock {\it {\JournalTitle{\apj}}} \textbf{\bibinfo{volume}{886}},
  \bibinfo{pages}{108} (\bibinfo{year}{2019}).

\bibitem{1998ApJ...500..525S}
\bibinfo{author}{{Schlegel}, D.~J.}, \bibinfo{author}{{Finkbeiner}, D.~P.} \&
  \bibinfo{author}{{Davis}, M.}
\newblock \bibinfo{journal}{\bibinfo{title}{{Maps of Dust Infrared Emission for
  Use in Estimation of Reddening and Cosmic Microwave Background Radiation
  Foregrounds}}}.
\newblock {\it {\JournalTitle{\apj}}} \textbf{\bibinfo{volume}{500}},
  \bibinfo{pages}{525--553} (\bibinfo{year}{1998}).

\bibitem{2011ApJ...737..103S}
\bibinfo{author}{{Schlafly}, E.~F.} \& \bibinfo{author}{{Finkbeiner}, D.~P.}
\newblock \bibinfo{journal}{\bibinfo{title}{{Measuring Reddening with Sloan
  Digital Sky Survey Stellar Spectra and Recalibrating SFD}}}.
\newblock {\it {\JournalTitle{\apj}}} \textbf{\bibinfo{volume}{737}},
  \bibinfo{pages}{103} (\bibinfo{year}{2011}).

\bibitem{kurtz1998}
\bibinfo{author}{{Kurtz}, M.~J.} \& \bibinfo{author}{{Mink}, D.~J.}
\newblock \bibinfo{journal}{\bibinfo{title}{{RVSAO 2.0: Digital Redshifts and
  Radial Velocities}}}.
\newblock {\it {\JournalTitle{\pasp}}} \textbf{\bibinfo{volume}{110}},
  \bibinfo{pages}{934--977} (\bibinfo{year}{1998}).

\bibitem{emcee}
\bibinfo{author}{{Foreman-Mackey}, D.}, \bibinfo{author}{{Hogg}, D.~W.},
  \bibinfo{author}{{Lang}, D.} \& \bibinfo{author}{{Goodman}, J.}
\newblock \bibinfo{journal}{\bibinfo{title}{{emcee: The MCMC Hammer}}}.
\newblock {\it {\JournalTitle{\pasp}}} \textbf{\bibinfo{volume}{125}},
  \bibinfo{pages}{306} (\bibinfo{year}{2013}).

\bibitem{bovy2015}
\bibinfo{author}{{Bovy}, J.}
\newblock \bibinfo{journal}{\bibinfo{title}{{galpy: A python Library for
  Galactic Dynamics}}}.
\newblock {\it {\JournalTitle{\apjs}}} \textbf{\bibinfo{volume}{216}},
  \bibinfo{pages}{29} (\bibinfo{year}{2015}).

\bibitem{bovy2013}
\bibinfo{author}{{Bovy}, J.} \& \bibinfo{author}{{Rix}, H.-W.}
\newblock \bibinfo{journal}{\bibinfo{title}{{A Direct Dynamical Measurement of
  the Milky Way's Disk Surface Density Profile, Disk Scale Length, and Dark
  Matter Profile at 4 kpc \&lt;\raisebox{-0.5ex}\textasciitilde R
  \&lt;\raisebox{-0.5ex}\textasciitilde 9 kpc}}}.
\newblock {\it {\JournalTitle{\apj}}} \textbf{\bibinfo{volume}{779}},
  \bibinfo{pages}{115} (\bibinfo{year}{2013}).

\bibitem{schonrich2010}
\bibinfo{author}{{Sch{\"o}nrich}, R.}, \bibinfo{author}{{Binney}, J.} \&
  \bibinfo{author}{{Dehnen}, W.}
\newblock \bibinfo{journal}{\bibinfo{title}{{Local kinematics and the local
  standard of rest}}}.
\newblock {\it {\JournalTitle{\mnras}}} \textbf{\bibinfo{volume}{403}},
  \bibinfo{pages}{1829--1833} (\bibinfo{year}{2010}).

\bibitem{schonrich2012}
\bibinfo{author}{{Sch{\"o}nrich}, R.}
\newblock \bibinfo{journal}{\bibinfo{title}{{Galactic rotation and solar motion
  from stellar kinematics}}}.
\newblock {\it {\JournalTitle{\mnras}}} \textbf{\bibinfo{volume}{427}},
  \bibinfo{pages}{274--287} (\bibinfo{year}{2012}).

\bibitem{pauli2006}
\bibinfo{author}{{Pauli}, E.~M.}, \bibinfo{author}{{Napiwotzki}, R.},
  \bibinfo{author}{{Heber}, U.}, \bibinfo{author}{{Altmann}, M.} \&
  \bibinfo{author}{{Odenkirchen}, M.}
\newblock \bibinfo{journal}{\bibinfo{title}{{3D kinematics of white dwarfs from
  the SPY project. II.}}}
\newblock {\it {\JournalTitle{\aap}}} \textbf{\bibinfo{volume}{447}},
  \bibinfo{pages}{173--184} (\bibinfo{year}{2006}).

\bibitem{claret2020}
\bibinfo{author}{{Claret}, A.} et~al.
\newblock \bibinfo{journal}{\bibinfo{title}{{Gravity and limb-darkening
  coefficients for compact stars: DA, DB, and DBA eclipsing white dwarfs}}}.
\newblock {\it {\JournalTitle{\aap}}} \textbf{\bibinfo{volume}{634}},
  \bibinfo{pages}{A93} (\bibinfo{year}{2020}).

\bibitem{lenz2014}
\bibinfo{author}{{Lenz}, P.} \& \bibinfo{author}{{Breger}, M.}
\newblock \bibinfo{title}{{Period04: Statistical analysis of large astronomical
  time series}} (\bibinfo{year}{2014}).

\bibitem{MESA1}
\bibinfo{author}{{Paxton}, B.} et~al.
\newblock \bibinfo{journal}{\bibinfo{title}{{Modules for Experiments in Stellar
  Astrophysics (MESA)}}}.
\newblock {\it {\JournalTitle{\apjs}}} \textbf{\bibinfo{volume}{192}},
  \bibinfo{pages}{3} (\bibinfo{year}{2011}).

\bibitem{MESA2}
\bibinfo{author}{{Paxton}, B.} et~al.
\newblock \bibinfo{journal}{\bibinfo{title}{{Modules for Experiments in Stellar
  Astrophysics (MESA): Planets, Oscillations, Rotation, and Massive Stars}}}.
\newblock {\it {\JournalTitle{\apjs}}} \textbf{\bibinfo{volume}{208}},
  \bibinfo{pages}{4} (\bibinfo{year}{2013}).

\bibitem{MESA3}
\bibinfo{author}{{Paxton}, B.} et~al.
\newblock \bibinfo{journal}{\bibinfo{title}{{Modules for Experiments in Stellar
  Astrophysics (MESA): Binaries, Pulsations, and Explosions}}}.
\newblock {\it {\JournalTitle{\apjs}}} \textbf{\bibinfo{volume}{220}},
  \bibinfo{pages}{15} (\bibinfo{year}{2015}).

\bibitem{MESA4}
\bibinfo{author}{{Paxton}, B.} et~al.
\newblock \bibinfo{journal}{\bibinfo{title}{{Modules for Experiments in Stellar
  Astrophysics (MESA): Convective Boundaries, Element Diffusion, and Massive
  Star Explosions}}}.
\newblock {\it {\JournalTitle{\apjs}}} \textbf{\bibinfo{volume}{234}},
  \bibinfo{pages}{34} (\bibinfo{year}{2018}).

\bibitem{MESA5}
\bibinfo{author}{{Paxton}, B.} et~al.
\newblock \bibinfo{journal}{\bibinfo{title}{{Modules for Experiments in Stellar
  Astrophysics (MESA): Pulsating Variable Stars, Rotation, Convective
  Boundaries, and Energy Conservation}}}.
\newblock {\it {\JournalTitle{\apjs}}} \textbf{\bibinfo{volume}{243}},
  \bibinfo{pages}{10} (\bibinfo{year}{2019}).

\bibitem{Ostrowski2020}
\bibinfo{author}{{Ostrowski}, J.}, \bibinfo{author}{{Baran}, A.~S.},
  \bibinfo{author}{{Sanjayan}, S.} \& \bibinfo{author}{{Sahoo}, S.~K.}
\newblock \bibinfo{journal}{\bibinfo{title}{{Evolutionary modelling of subdwarf
  B stars using MESA with the predictive mixing and convective pre-mixing
  schemes}}}.
\newblock {\it {\JournalTitle{\mnras}}} \textbf{\bibinfo{volume}{503}},
  \bibinfo{pages}{4646--4661} (\bibinfo{year}{2021}).

\bibitem{jager1988}
\bibinfo{author}{{de Jager}, C.}, \bibinfo{author}{{Nieuwenhuijzen}, H.} \&
  \bibinfo{author}{{van der Hucht}, K.~A.}
\newblock \bibinfo{journal}{\bibinfo{title}{{Mass loss rates in the
  Hertzsprung-Russell diagram.}}}
\newblock {\it {\JournalTitle{\aaps}}} \textbf{\bibinfo{volume}{72}},
  \bibinfo{pages}{259--289} (\bibinfo{year}{1988}).

\bibitem{Krticka2019}
\bibinfo{author}{{Krti{\v{c}}ka}, J.} et~al.
\newblock \bibinfo{journal}{\bibinfo{title}{{Hot subdwarf wind models with
  accurate abundances. I. Hydrogen dominated stars HD 49798 and BD+18º2647}}}.
\newblock {\it {\JournalTitle{\aap}}} \textbf{\bibinfo{volume}{631}},
  \bibinfo{pages}{A75} (\bibinfo{year}{2019}).

\bibitem{Zenati2019}
\bibinfo{author}{{Zenati}, Y.}, \bibinfo{author}{{Toonen}, S.} \&
  \bibinfo{author}{{Perets}, H.~B.}
\newblock \bibinfo{journal}{\bibinfo{title}{{Formation and evolution of hybrid
  He-CO white dwarfs and their properties}}}.
\newblock {\it {\JournalTitle{\mnras}}} \textbf{\bibinfo{volume}{482}},
  \bibinfo{pages}{1135--1142} (\bibinfo{year}{2019}).

\bibitem{nomoto1980}
\bibinfo{author}{{Nomoto}, K.}
\newblock \bibinfo{title}{{Supernova explosions in accreting whiteddwarfs and
  Type I supernovae}}.
\newblock In \bibinfo{editor}{{Wheeler}, J.~C.} (ed.)
  {\bibinfo{booktitle}{Texas Workshop on Type I Supernovae}},
  \bibinfo{pages}{164--181} (\bibinfo{year}{1980}).

\bibitem{livne1990}
\bibinfo{author}{{Livne}, E.}
\newblock \bibinfo{journal}{\bibinfo{title}{{Successive Detonations in
  Accreting White Dwarfs as an Alternative Mechanism for Type I Supernovae}}}.
\newblock {\it {\JournalTitle{\apjl}}} \textbf{\bibinfo{volume}{354}},
  \bibinfo{pages}{L53} (\bibinfo{year}{1990}).

\bibitem{shen2014}
\bibinfo{author}{{Shen}, K.~J.} \& \bibinfo{author}{{Bildsten}, L.}
\newblock \bibinfo{journal}{\bibinfo{title}{{The Ignition of Carbon Detonations
  via Converging Shock Waves in White Dwarfs}}}.
\newblock {\it {\JournalTitle{\apj}}} \textbf{\bibinfo{volume}{785}},
  \bibinfo{pages}{61} (\bibinfo{year}{2014}).

\bibitem{pakmor2011}
\bibinfo{author}{{Pakmor}, R.}, \bibinfo{author}{{Hachinger}, S.},
  \bibinfo{author}{{R{\"o}pke}, F.~K.} \& \bibinfo{author}{{Hillebrand t}, W.}
\newblock \bibinfo{journal}{\bibinfo{title}{{Violent mergers of nearly
  equal-mass white dwarf as progenitors of subluminous Type Ia supernovae}}}.
\newblock {\it {\JournalTitle{\aap}}} \textbf{\bibinfo{volume}{528}},
  \bibinfo{pages}{A117} (\bibinfo{year}{2011}).

\bibitem{pakmor2012}
\bibinfo{author}{{Pakmor}, R.} et~al.
\newblock \bibinfo{journal}{\bibinfo{title}{{Normal Type Ia Supernovae from
  Violent Mergers of White Dwarf Binaries}}}.
\newblock {\it {\JournalTitle{\apjl}}} \textbf{\bibinfo{volume}{747}},
  \bibinfo{pages}{L10} (\bibinfo{year}{2012}).

\bibitem{röpke2012}
\bibinfo{author}{{R{\"o}pke}, F.~K.} et~al.
\newblock \bibinfo{journal}{\bibinfo{title}{{Constraining Type Ia Supernova
  Models: SN 2011fe as a Test Case}}}.
\newblock {\it {\JournalTitle{\apjl}}} \textbf{\bibinfo{volume}{750}},
  \bibinfo{pages}{L19} (\bibinfo{year}{2012}).

\bibitem{sato2016}
\bibinfo{author}{{Sato}, Y.} et~al.
\newblock \bibinfo{journal}{\bibinfo{title}{{The Critical Mass Ratio of Double
  White Dwarf Binaries for Violent Merger-induced Type Ia Supernova
  Explosions}}}.
\newblock {\it {\JournalTitle{\apj}}} \textbf{\bibinfo{volume}{821}},
  \bibinfo{pages}{67} (\bibinfo{year}{2016}).

\bibitem{li2003}
\bibinfo{author}{{Li}, W.} et~al.
\newblock \bibinfo{journal}{\bibinfo{title}{{SN 2002cx: The Most Peculiar Known
  Type Ia Supernova}}}.
\newblock {\it {\JournalTitle{\pasp}}} \textbf{\bibinfo{volume}{115}},
  \bibinfo{pages}{453--473} (\bibinfo{year}{2003}).

\bibitem{foley2013}
\bibinfo{author}{{Foley}, R.~J.} et~al.
\newblock \bibinfo{journal}{\bibinfo{title}{{Type Iax Supernovae: A New Class
  of Stellar Explosion}}}.
\newblock {\it {\JournalTitle{\apj}}} \textbf{\bibinfo{volume}{767}},
  \bibinfo{pages}{57} (\bibinfo{year}{2013}).

\bibitem{wang2013}
\bibinfo{author}{{Wang}, B.}, \bibinfo{author}{{Justham}, S.} \&
  \bibinfo{author}{{Han}, Z.}
\newblock \bibinfo{journal}{\bibinfo{title}{{Producing Type Iax supernovae from
  a specific class of helium-ignited WD explosions}}}.
\newblock {\it {\JournalTitle{\aap}}} \textbf{\bibinfo{volume}{559}},
  \bibinfo{pages}{A94} (\bibinfo{year}{2013}).

\bibitem{han2002}
\bibinfo{author}{{Han}, Z.}, \bibinfo{author}{{Podsiadlowski}, P.},
  \bibinfo{author}{{Maxted}, P.~F.~L.}, \bibinfo{author}{{Marsh}, T.~R.} \&
  \bibinfo{author}{{Ivanova}, N.}
\newblock \bibinfo{journal}{\bibinfo{title}{{The origin of subdwarf B stars -
  I. The formation channels}}}.
\newblock {\it {\JournalTitle{\mnras}}} \textbf{\bibinfo{volume}{336}},
  \bibinfo{pages}{449--466} (\bibinfo{year}{2002}).

\bibitem{weidemann2000}
\bibinfo{author}{{Weidemann}, V.}
\newblock \bibinfo{journal}{\bibinfo{title}{{Revision of the initial-to-final
  mass relation}}}.
\newblock {\it {\JournalTitle{\aap}}} \textbf{\bibinfo{volume}{363}},
  \bibinfo{pages}{647--656} (\bibinfo{year}{2000}).

\bibitem{astropy:2013}
\bibinfo{author}{{Astropy Collaboration}} et~al.
\newblock \bibinfo{journal}{\bibinfo{title}{{Astropy: A community Python
  package for astronomy}}}.
\newblock {\it {\JournalTitle{\aap}}} \textbf{\bibinfo{volume}{558}},
  \bibinfo{pages}{A33} (\bibinfo{year}{2013}).

\bibitem{astropy:2018}
\bibinfo{author}{{Price-Whelan}, A.~M.} et~al.
\newblock \bibinfo{journal}{\bibinfo{title}{{The Astropy Project: Building an
  Open-science Project and Status of the v2.0 Core Package}}}.
\newblock {\it {\JournalTitle{\aj}}} \textbf{\bibinfo{volume}{156}},
  \bibinfo{pages}{123} (\bibinfo{year}{2018}).

\bibitem{2015AAS...22533616H}
\bibinfo{author}{{Henden}, A.~A.}, \bibinfo{author}{{Levine}, S.},
  \bibinfo{author}{{Terrell}, D.} \& \bibinfo{author}{{Welch}, D.~L.}
\newblock \bibinfo{title}{{APASS - The Latest Data Release}}.
\newblock In {\bibinfo{booktitle}{American Astronomical Society Meeting
  Abstracts \#225}}, vol. \bibinfo{volume}{225} of {\bibinfo{series}{American
  Astronomical Society Meeting Abstracts}}, \bibinfo{pages}{336.16}
  (\bibinfo{year}{2015}).

\bibitem{2017yCat.2349....0C}
\bibinfo{author}{{Chambers}, K.~C.} et~al.
\newblock \bibinfo{journal}{\bibinfo{title}{{VizieR Online Data Catalog: The
  Pan-STARRS release 1 (PS1) Survey - DR1 (Chambers+, 2016)}}}.
\newblock {\it {\JournalTitle{VizieR Online Data Catalog}}}
  \textbf{\bibinfo{volume}{2349}} (\bibinfo{year}{2017}).

\bibitem{2006AJ....131.1163S}
\bibinfo{author}{{Skrutskie}, M.~F.} et~al.
\newblock \bibinfo{journal}{\bibinfo{title}{{The Two Micron All Sky Survey
  (2MASS)}}}.
\newblock {\it {\JournalTitle{\aj}}} \textbf{\bibinfo{volume}{131}},
  \bibinfo{pages}{1163--1183} (\bibinfo{year}{2006}).

\bibitem{2018A&A...616A...4E}
\bibinfo{author}{{Evans}, D.~W.} et~al.
\newblock \bibinfo{journal}{\bibinfo{title}{{Gaia Data Release 2. Photometric
  content and validation}}}.
\newblock {\it {\JournalTitle{\aap}}} \textbf{\bibinfo{volume}{616}},
  \bibinfo{pages}{A4} (\bibinfo{year}{2018}).

\bibitem{2018A&A...619A.180M}
\bibinfo{author}{{Ma{\'\i}z Apell{\'a}niz}, J.} \& \bibinfo{author}{{Weiler},
  M.}
\newblock \bibinfo{journal}{\bibinfo{title}{{Reanalysis of the Gaia Data
  Release 2 photometric sensitivity curves using HST/STIS spectrophotometry}}}.
\newblock {\it {\JournalTitle{\aap}}} \textbf{\bibinfo{volume}{619}},
  \bibinfo{pages}{A180} (\bibinfo{year}{2018}).

\bibitem{2014yCat.2328....0C}
\bibinfo{author}{{Cutri}, R.~M.} \& \bibinfo{author}{{et al.}}
\newblock \bibinfo{journal}{\bibinfo{title}{{VizieR Online Data Catalog:
  AllWISE Data Release (Cutri+ 2013)}}}.
\newblock {\it {\JournalTitle{VizieR Online Data Catalog}}}
  \bibinfo{pages}{II/328} (\bibinfo{year}{2014}).

\bibitem{2019ApJS..240...30S}
\bibinfo{author}{{Schlafly}, E.~F.}, \bibinfo{author}{{Meisner}, A.~M.} \&
  \bibinfo{author}{{Green}, G.~M.}
\newblock \bibinfo{journal}{\bibinfo{title}{{The unWISE Catalog: Two Billion
  Infrared Sources from Five Years of WISE Imaging}}}.
\newblock {\it {\JournalTitle{\apjs}}} \textbf{\bibinfo{volume}{240}},
  \bibinfo{pages}{30} (\bibinfo{year}{2019}).

\bibitem{corner}
\bibinfo{author}{Foreman-Mackey, D.}
\newblock \bibinfo{journal}{\bibinfo{title}{corner.py: Scatterplot matrices in
  python}}.
\newblock {\it {\JournalTitle{The Journal of Open Source Software}}}
  \textbf{\bibinfo{volume}{1}}, \bibinfo{pages}{24} (\bibinfo{year}{2016}).

\end{thebibliography}

Correspondence and requests for materials should be addressed to Dr Ingrid Pelisoli (ingrid.pelisoli@warwick.ac.uk).

\section*{Acknowledgements} 

IP and VS were partially funded by the Deutsche Forschungsgemeinschaft (DFG) under grant GE2506/12-1. IP also acknowledges funding by the UK's Science and Technology Facilities Council (STFC), grant ST/T000406/1. PN gratefully acknowledges funding provided by the Max Planck society. AI
acknowledges funding by the DFG through grant HE1356/71-1. DS was supported by the DFG under grants HE 1356/70-1 and IR 190/1-1. BB acknowledges support from NASA under TESS Guest Investigator program grant 80NSSC19K1720. TK acknowledges support by the US National Science Foundation through grant No. NSF PHY-1748958. We thank T. R. Marsh for enlightening discussions and for providing an MCMC wrapper to be used with {\sc lcurve}. We are grateful to Andrzej S. Baran and David Jones for providing helpful comments to an earlier version of this manuscript.
     
This research made extensive use of Astropy (http://www.astropy.org) a community-developed core Python package for Astronomy \cite{astropy:2013, astropy:2018}
     
This paper includes data collected by the TESS mission. Funding for the TESS mission is provided by the NASA Explorer Program.

This work has made use of data from the European Space Agency (ESA) mission
{\it Gaia} (\url{https://www.cosmos.esa.int/gaia}), processed by the {\it Gaia}
Data Processing and Analysis Consortium (DPAC,
\url{https://www.cosmos.esa.int/web/gaia/dpac/consortium}). Funding for the DPAC
has been provided by national institutions, in particular the institutions
participating in the {\it Gaia} Multilateral Agreement.

Some of the data presented herein were obtained at the W.M. Keck Observatory, which is operated as a scientific partnership among the California Institute of Technology, the University of California and the National Aeronautics and Space Administration. The Observatory was made possible by the generous financial support of the W.M. Keck Foundation. The authors wish to recognise and acknowledge the very significant cultural role and reverence that the summit of Mauna Kea has always had within the indigenous Hawaiian community. We are most fortunate to have the opportunity to conduct observations from this mountain.

\section*{Author contributions}

All authors contributed to the work presented in this paper. IP carried out the radial velocity estimates and fitting, the light curve fitting, and led the writing of the manuscript. PN calculated the evolution of the system. SG and UH performed the spectral fitting. TK did the spectroscopic reduction and cross-checked the light curve fitting. DS and UH performed the SED fitting. AI wrote the SED fitting tool and calculated the spectral models used for SED and spectral fitting. AB calculated the Galactic orbit of the system. JvR performed the spectroscopic observations and contributed to the light curve fit. VS and BNB contributed to the analysis of the light curve. All authors reviewed the manuscript.

\section*{Competing interests statement}
The authors declare no competing interests.

%\section*{Figure Legends and Tables}

\section*{Supplementary Information}

\begin{suptable}[h!]
\caption{{\bf Identified short periods of the system.} The 33 frequencies above a 5-$\sigma$ detection level, calculated using the average amplitude of the Fourier transform, that were identified and removed from the light curve prior to fitting a binary model. Uncertainties were obtained with 500 Monte Carlo runs in {\sc Period}04\cite{lenz2014} and are given on the last significant digits, e.g. $3371.859(7) = 3371.859\pm0.007$.}    
\label{table:pmods}
\centering 
\begin{tabular}{c c c}
\hline\hline 
ID & Frequency ($\mu$Hz) & Amplitude (ppt) \\
\hline 
f1 & 3371.859(7) & 2.34(7) \\
f2 & 3035.503(7) & 2.23(7) \\
f3 & 3203.680(11) & 1.45(7) \\
f4 & 3050.4(8) & 1.25(28) \\
f5 & 3219(19) & 1.03(39) \\
f6 & 3431.656(16) & 1.00(7) \\
f7 & 3263.410(19) & 0.87(7) \\
f8 & 2882.22(29) & 0.89(8) \\
f9 & 3057(27) & 0.88(39) \\
f10 & 3095(37) & 1.1(7) \\
f11 & 3540.0(9) & 0.86(15) \\
f12 & 2926.9(22) & 0.84(20) \\
f13 & 3387(14) & 0.81(34) \\
f14 & 3385.7(1.9) & 0.78(14) \\
f15 & 3049.4(9) & 0.75(27) \\
f16 & 4127(15) & 0.71(31) \\
f17 & 3212.832(24) & 0.71(7) \\
f18 & 2714(37) & 0.68(29) \\
f19 & 3429.249(23) & 0.68(7) \\
f20 & 2888(169) & 0.65(27) \\
f21 & 3225.097(25) & 0.65(7) \\
f22 & 2867.282(26) & 0.63(7) \\
f23 & 3668.8(3.2) & 0.55(12) \\
f24 & 3393(51) &  0.51(22) \\
f25 & 2759(13) & 0.49(16) \\
f26 & 3768(22) & 0.47(21) \\
f27 & 3095(66) & 0.63(74) \\
f28 & 3779.122(36) & 0.42(7) \\
f29 & 2881.204(38) & 0.43(8) \\
f30 & 2720.577(38) & 0.41(7) \\
f31 & 3261.03(34) & 0.41(8) \\
f32 & 2208.5(5) & 0.39(8) \\
f33 & 2544.78(30) & 0.38(7) \\
\hline
\end{tabular}
\end{suptable}

\begin{suppfigure*}[h!]
\centering
\includegraphics[width=\hsize]{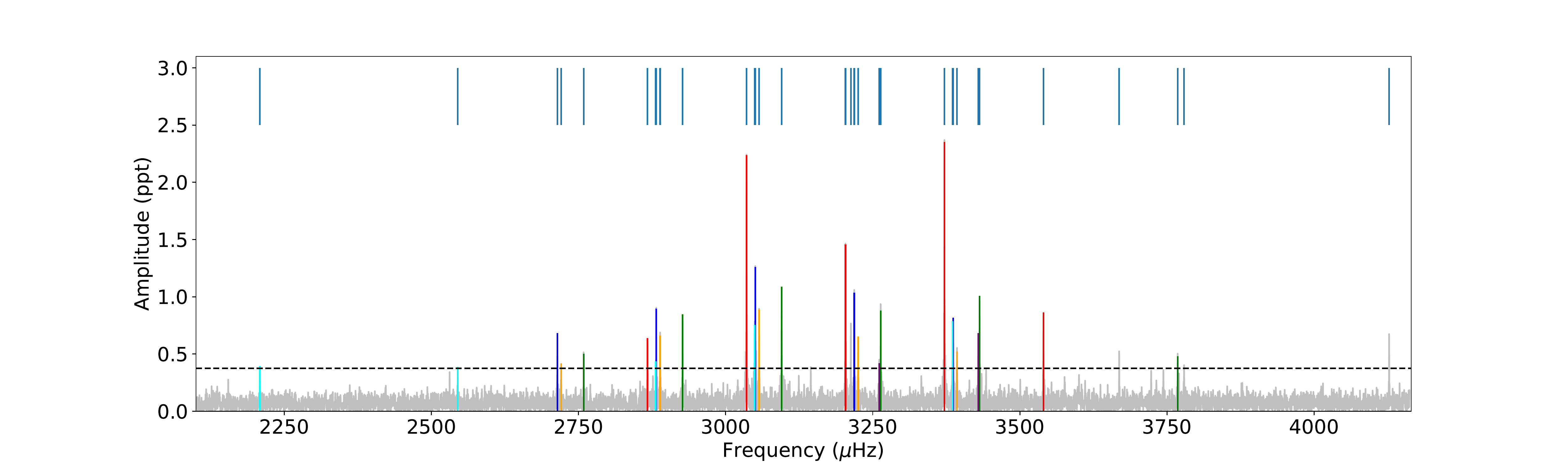}
\caption{{\bf Pulsations of the hot subdwarf.} Fourier transform showing the region with identified short-period peaks (marked by the black vertical lines). The three strongest peaks have a separation of $168.179\pm0.013~\mu$Hz, in agreement with the orbital period ($168.183~~\mu$Hz). Other peaks that are separated by multiples of the orbital period are marked in the same colour. The dashed vertical line was our final detection threshold. Periods above the detection limit were subtracted prior to the light curve modelling.}
\label{pmodes}
\end{suppfigure*}

\begin{suppfigure*}
   \centering
   \includegraphics[width=\hsize]{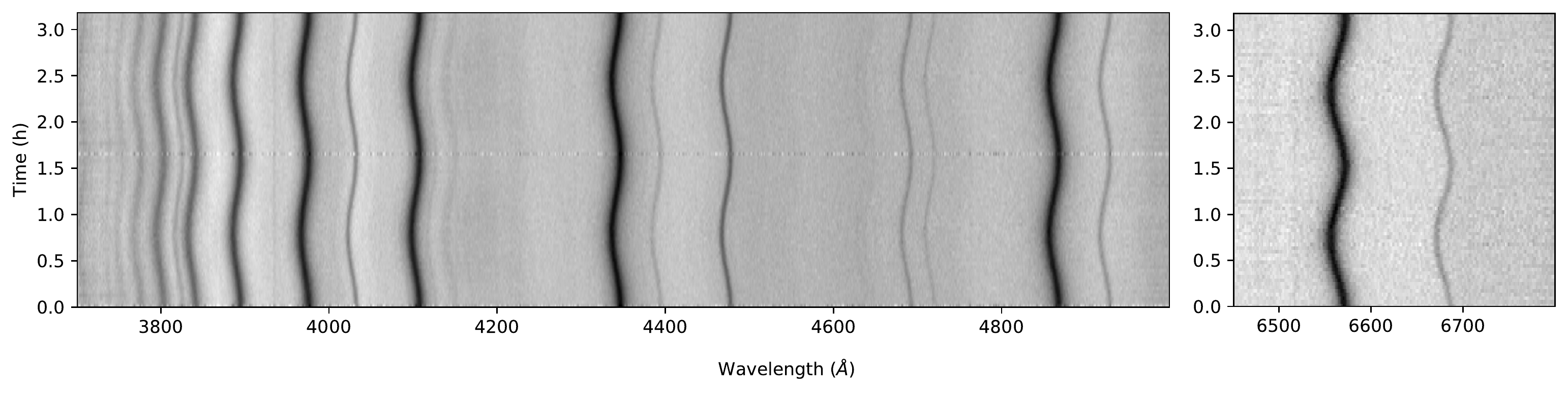}
      \caption{{\bf Trailed spectra of the exposures obtained with the Palomar 200-inch telescope.} We plot the same spectra twice to cover two cycles and aid visualisation of the radial velocity changes.}
         \label{trail_spec}
   \end{suppfigure*}
   
\begin{suppfigure*}
    \centering
    
    \begin{minipage}{.5\textwidth}
    \centering
    \includegraphics[width=\hsize]{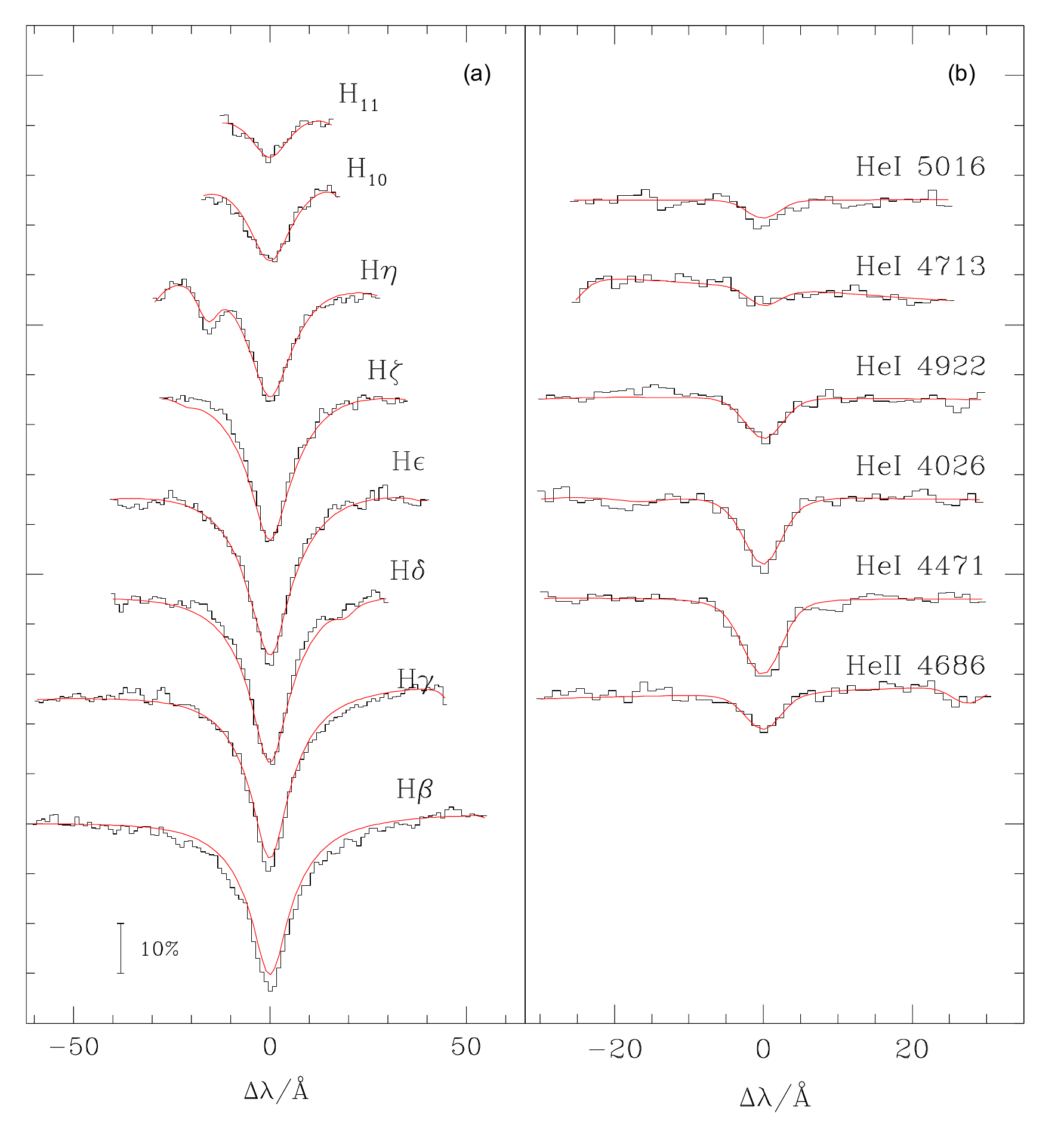}
    \end{minipage}%
    \begin{minipage}{.5\textwidth}
    \centering
    \includegraphics[width=\hsize]{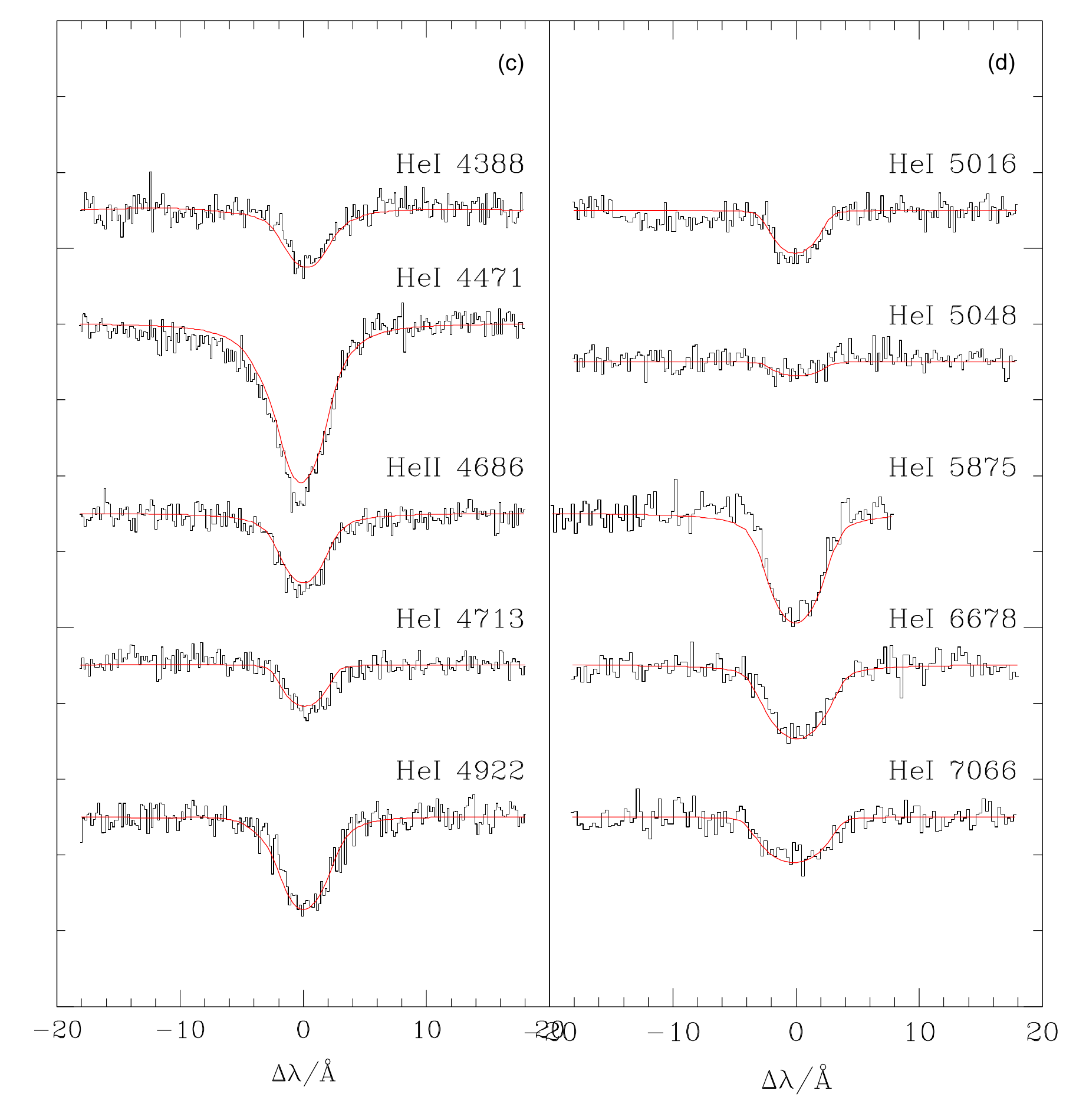}
    \end{minipage}

        \caption{{\bf Spectroscopic fit to individual DBSP and ESI spectra.} Panels (a) and (b) show, respectively, the fit to the hydrogen and helium lines of a DBSP spectrum. Panels (c) and (d) show the fit to one of the ESI spectra. The observed spectra are shown in black, and the fitted models are shown in red. The respective orbital phases of each spectrum are given in panels b and c. Only the statistical uncertainties are quoted in this figure.}
    \label{spec_fit}
\end{suppfigure*}

\begin{suppfigure}
    \centering
  \includegraphics[width=1\columnwidth]{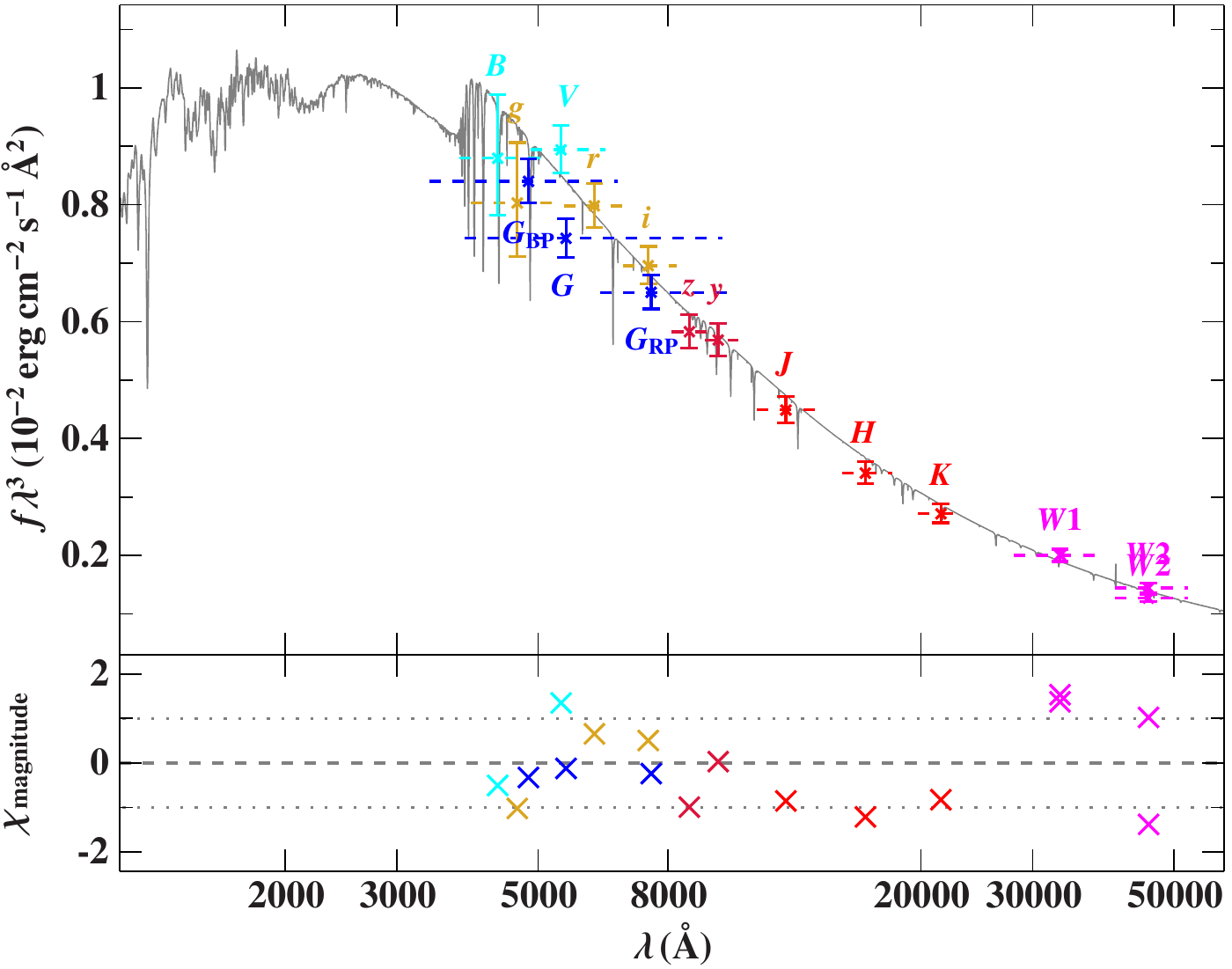}
    \caption{\label{fig:photometry_sed_single} {\bf Comparison of synthetic and
      observed photometry.} \textit{Top panel:} SED of HD~265435 with filter-averaged fluxes (times wavelength to the power of three) converted from observed magnitudes. 
      The approximate width of the respective filters (widths at 10\% of maximum) are shown by dashed lines. Error-bars represend the one-sigma uncertainties. The best-fitting model SED, smoothed
      to a spectral resolution of 6\,{\AA}, is shown in grey. 
      \textit{Bottom panel:} Residuals $\chi$, i.e., the difference between synthetic and observed magnitudes divided by the corresponding uncertainties.
      The different photometric systems are displayed in the following colours: 
     golden for APASS-griz\cite{2015AAS...22533616H}, red for  
     PAN-STARRS\cite{2017yCat.2349....0C} and 2MASS\cite{2006AJ....131.1163S}, blue for
     APASS-Johnson\cite{2015AAS...22533616H} and
      {\it Gaia} \cite{2018A&A...616A...4E, 2018A&A...619A.180M}, and magenta for
      WISE \cite{2014yCat.2328....0C, 2019ApJS..240...30S}.
     }
\end{suppfigure} 

\clearpage

\begin{suptable}[h!]
\caption{{\bf Blue-channel radial velocities.} Baricentric Julian Date (BJD), radial velocities, and  one-sigma uncertainties for spectra taken with the DBSP blue arm.} % title of Table
\label{rv_values_b}      % is used to refer this table in the text
\centering                          % used for centering table
\begin{tabular}{c c c}        % centered columns (4 columns)
\hline\hline                 % inserts double horizontal lines
BJD & $RV$ (km/s) & $\sigma_{RV}$ (km/s) \\    % table heading 
\hline                        % inserts single horizontal line
  2458909.63734 &   337.03 &  6.21 \\
  2458909.63898 &   342.66 &  6.91 \\
  2458909.64062 &   330.93 &  6.63 \\
  2458909.64226 &   318.61 &  6.59 \\
  2458909.64390 &   297.84 &  6.81 \\
  2458909.64554 &   277.98 &  5.59 \\
  2458909.64718 &   239.58 &  6.20 \\
  2458909.64881 &   204.68 &  5.80 \\
  2458909.65045 &   163.12 &  6.48 \\
  2458909.65209 &   109.46 &  6.31 \\
  2458909.65373 &    61.75 &  6.07 \\
  2458909.65537 &     3.07 &  5.32 \\
  2458909.65701 &   -38.98 &  5.97 \\
  2458909.65865 &   -87.83 &  6.45 \\
  2458909.66029 &  -137.82 &  6.15 \\
  2458909.66193 &  -190.80 &  5.25 \\
  2458909.66357 &  -225.72 &  5.70 \\
  2458909.66521 &  -258.85 &  5.75 \\
  2458909.66685 &  -288.02 &  6.07 \\
  2458909.66849 &  -307.18 &  5.90 \\
  2458909.67013 &  -326.84 &  6.70 \\
  2458909.67177 &  -334.37 &  6.46 \\
  2458909.67341 &  -335.73 &  6.89 \\
  2458909.67505 &  -326.83 &  6.31 \\
  2458909.67669 &  -313.77 &  6.65 \\
  2458909.67833 &  -293.57 &  5.47 \\
  2458909.67997 &  -261.68 &  6.07 \\
  2458909.68161 &  -230.62 &  5.72 \\
  2458909.68325 &  -189.10 &  5.74 \\
  2458909.68489 &  -147.20 &  5.69 \\
  2458909.68653 &   -95.98 &  5.30 \\
  2458909.68817 &   -50.23 &  5.56 \\
  2458909.68981 &     5.97 &  5.83 \\
  2458909.69145 &    49.20 &  5.99 \\
  2458909.69309 &   107.42 &  5.66 \\
  2458909.69473 &   149.59 &  5.75 \\
  2458909.69637 &   198.87 &  6.33 \\
  2458909.69800 &   236.96 &  6.15 \\
  2458909.69964 &   280.80 &  5.81 \\
  2458909.70128 &   306.14 &  6.61 \\
  2458909.70292 &   329.57 &  6.62 \\
  2458909.70456 &   343.37 &  5.99 \\
  2458909.70555 &   348.70 & 13.38 \\
  \hline                                   %inserts single line
\end{tabular}
\end{suptable}

\begin{suptable}[h!]
\caption{{\bf Red-channel radial velocities.} Baricentric Julian Date (BJD), radial velocities, and  one-sigma uncertainties for spectra taken with the DBSP red arm.} % title of Table
\label{rv_values_r}      % is used to refer this table in the text
\centering                          % used for centering table
\begin{tabular}{c c c}        % centered columns (4 columns)
\hline\hline                 % inserts double horizontal lines
BJD & $RV$ (km/s) & $\sigma_{RV}$ (km/s) \\    % table heading 
\hline                        % inserts single horizontal line
  2458909.63723 &   370.11 & 15.86 \\
  2458909.63889 &   398.05 & 14.45 \\
  2458909.64054 &   390.82 & 14.35 \\
  2458909.64220 &   342.00 & 11.94 \\
  2458909.64386 &   319.98 &  8.81 \\
  2458909.64551 &   288.98 &  9.11 \\
  2458909.64717 &   257.57 & 12.85 \\
  2458909.64883 &   238.55 &  9.51 \\
  2458909.65048 &   165.75 & 13.11 \\
  2458909.65214 &   126.50 & 11.93 \\
  2458909.65380 &    60.34 & 13.94 \\
  2458909.65545 &     8.46 & 12.71 \\
  2458909.65711 &   -43.24 & 13.29 \\
  2458909.65877 &   -63.40 & 14.26 \\
  2458909.66042 &  -115.31 &  8.95 \\
  2458909.66208 &  -165.17 & 11.84 \\
  2458909.66374 &  -218.92 & 11.69 \\
  2458909.66539 &  -251.66 & 14.47 \\
  2458909.66705 &  -315.21 &  8.73 \\
  2458909.66871 &  -297.70 & 18.51 \\
  2458909.67036 &  -352.65 & 21.30 \\
  2458909.67202 &  -297.57 & 11.92 \\
  2458909.67368 &  -329.15 & 23.94 \\
  2458909.67533 &  -298.43 & 21.41 \\
  2458909.67699 &  -304.48 & 16.63 \\
  2458909.67865 &  -270.85 & 11.25 \\
  2458909.68030 &  -256.84 & 14.66 \\
  2458909.68196 &  -212.91 & 12.39 \\
  2458909.68362 &  -168.86 & 12.36 \\
  2458909.68527 &  -114.14 & 10.04 \\
  2458909.68693 &   -82.01 & 10.88 \\
  2458909.68859 &   -52.29 & 11.89 \\
  2458909.69024 &    36.37 &  8.93 \\
  2458909.69190 &    78.97 & 10.56 \\
  2458909.69355 &   148.34 & 13.11 \\
  2458909.69521 &   158.96 & 12.96 \\
  2458909.69687 &   208.76 & 10.13 \\
  2458909.69852 &   229.36 & 13.41 \\
  2458909.70018 &   292.78 & 11.64 \\
  2458909.70184 &   299.48 & 12.12 \\
  2458909.70349 &   342.46 &  9.54 \\
  \hline                                   %inserts single line
\end{tabular}
\end{suptable}

\clearpage

\begin{suppfigure}
\centering
\includegraphics[width=0.75\hsize]{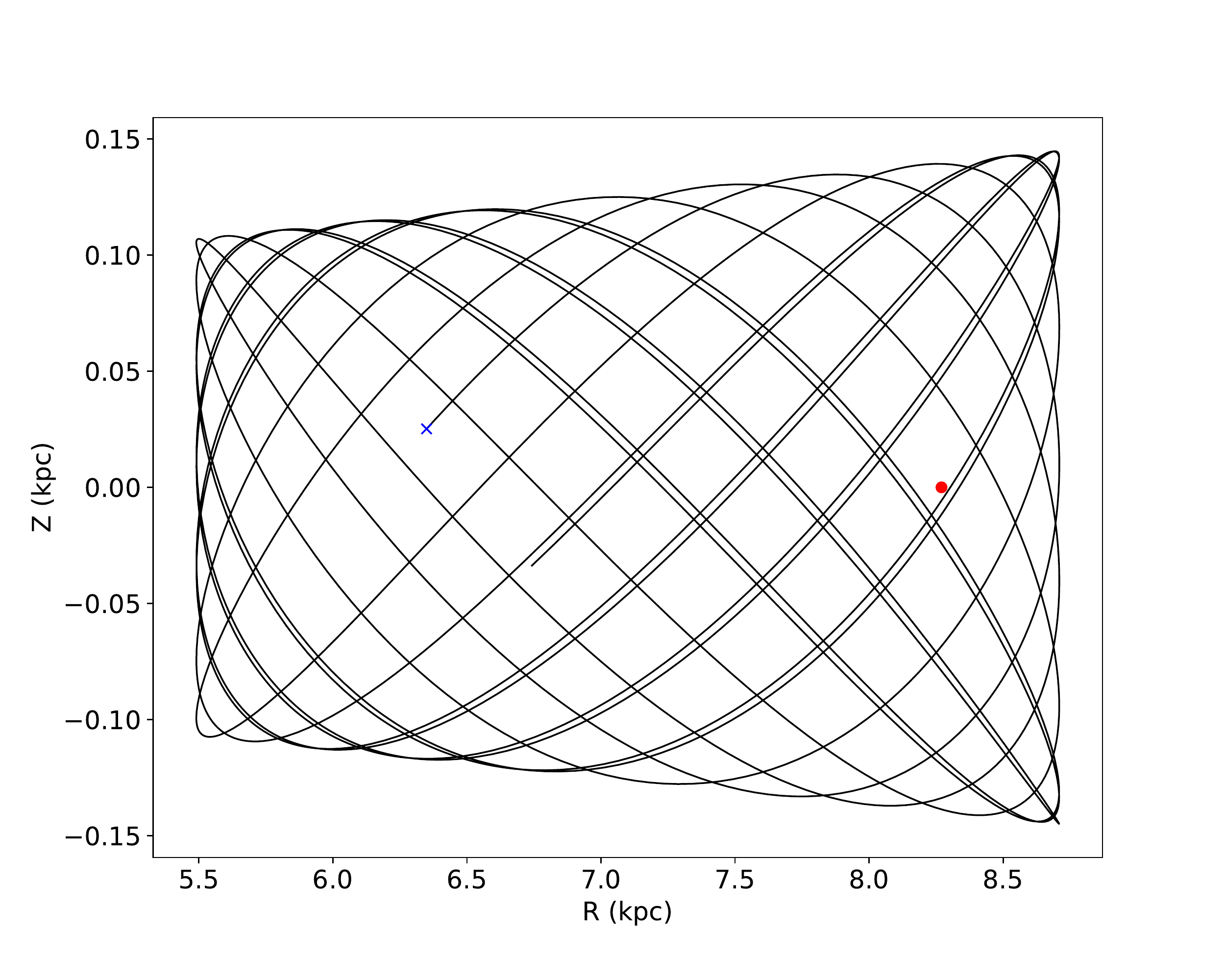}
\caption{{\bf Galactocentric orbit for HD~265435.} For illustrative purposes, we used a lookback time of 1~Gyr. We show the height above the disk ($Z$) as a function of Galactocentric distance $R = \sqrt{X^2 + Y^2}$. The current position of the Sun is indicated with a red dot for reference, and the position of HD~265435 is marked with a blue cross.}
\label{orbit}
\end{suppfigure}

\begin{suppfigure}
\centering
\includegraphics[width=0.75\hsize]{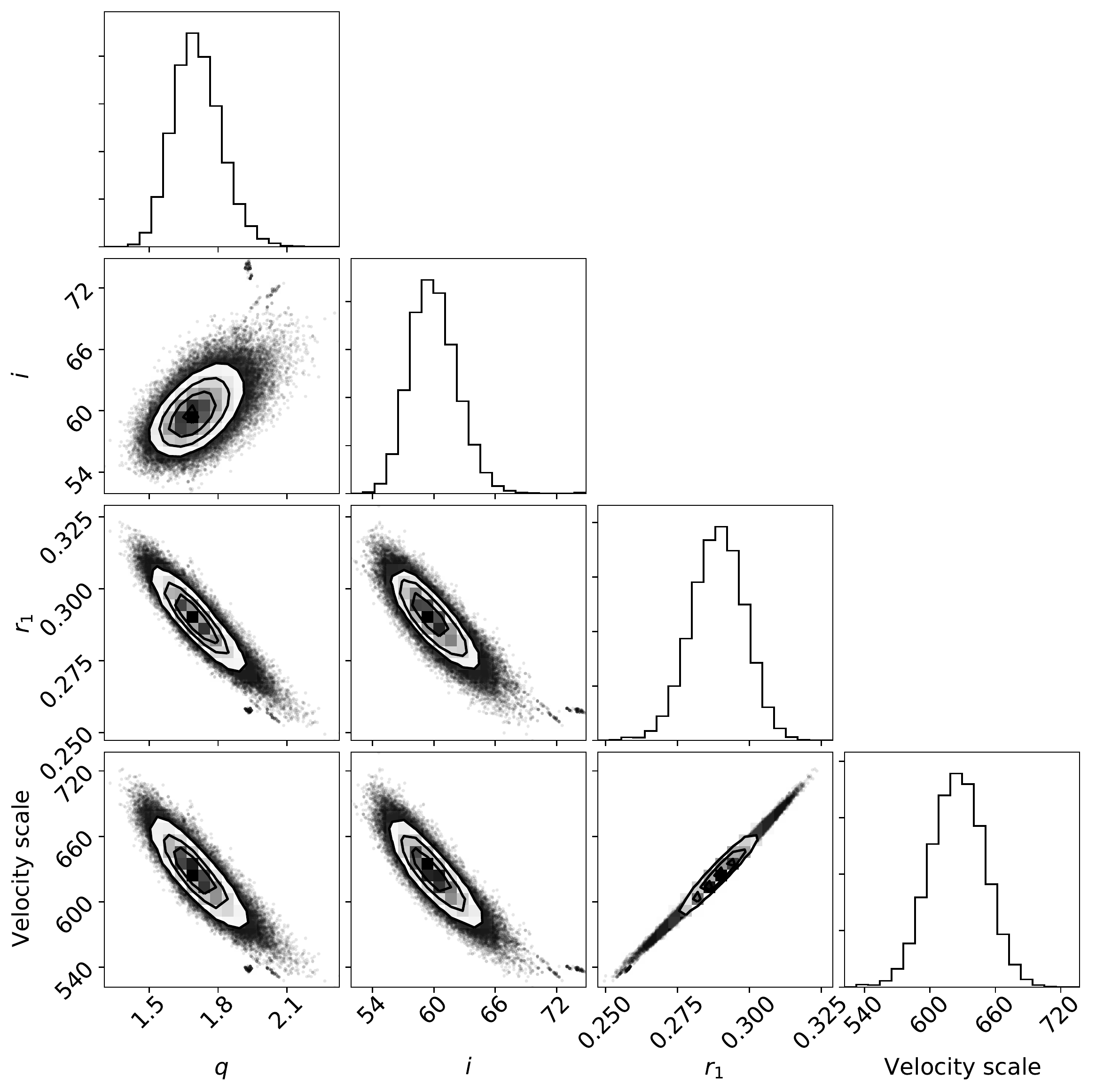}
\caption{{\bf Corner plot\cite{corner} for light curve MCMC fit}. Posterior probability distributions for the parameters directly derived from the MCMC fit to the TESS light curve.}
\label{corner}
\end{suppfigure}

%TC:endignore

\end{document}